\documentclass[useAMS,usenatbib]{mn2e}
\usepackage{graphicx}

\title[AO parameters-wind connection]{Adaptive Optics Parameters connection to wind speed at the Teide Observatory}

\author[Garc\'{\i}a-Lorenzo et al.]{B. Garc\'{\i}a-Lorenzo$^{1}$\thanks{E-mail:
bgarcia@iac.es}, A. Eff-Darwich$^{2}$, J. J. Fuensalida$^{1}$, \& J. Castro-Almaz\'an$^{1}$\\
1 Instituto de Astrof\'{\i}sica de Canarias, C/Via Lactea S/N, 38305-La Laguna, Tenerife, Spain \\
2 Dept. Edafolog\'{\i}a y Geolog\'{\i}a, Universidad de La Laguna, C/ Astrof\'{\i}sico Francisco S\'anchez, E-38205 Tenerife, Spain}

\begin{document}

%\date{Accepted ..... Received .....; in original form .....}

\pagerange{\pageref{firstpage}--\pageref{lastpage}} \pubyear{2004}

\maketitle

\label{firstpage}

\begin{abstract} 
Current projects for large telescopes demand a proper knowledge of atmospheric turbulence to design 
efficient adaptive optics systems in order to reach large Strehl ratios. However, the proper characterization 
of the turbulence above a particular site requires long-term monitoring. Due to the lack of long-term 
information on turbulence, high-altitude winds (in particular winds at the 200 mbar pressure level) were 
proposed as a parameter for estimating the total turbulence at a particular site, with the advantage of 
records of winds going back several decades. We present the first complete study of atmospheric adaptive optics 
parameters above the Teide Observatory (Canary Islands, Spain) in relation to wind speed. On-site 
measurements of $C_N^2(h)$ profiles (more than 20200 turbulence profiles) from G-SCIDAR observations 
and wind vertical profiles from balloons have been used to calculate the seeing, the isoplanatic angle 
and the coherence time. The connection of these parameters to wind speeds at ground and 200 mbar pressure 
level are shown and discussed. Our results confirm the well-known high quality of the Canary Islands astronomical observatories. 

%{\bf Total number of individual profiles used for this study: 20232} 

\end{abstract}


\section{Introduction}

The presence of optical turbulence in the Earth's atmosphere drastically affects ground-based astronomical observations. The wavefront of the light
 coming from astronomical objects is distorted when passing through the
 turbulence layers,  the wavefront being aleatory when reaching the entrance
 pupil of telescopes. The result is a degradation of the angular resolution
 of  ground-based astronomical instruments. Several techniques have been
 developed to compensate for the effects of the atmosphere on astronomical
 images trying to reach the diffraction limit,  the most
 popular being  adaptive optics (AO hereafter) systems. The larger the telescope diameter, the more difficult 
 the proper correction of the atmospheric turbulence becomes. The excellent image quality requirements of current large and future extremely large
 telescopes needs the design of adaptive optic systems with the capacity 
of adaptability to the prevailing turbulence conditions at the observing site.
A proper knowledge of the statistical behaviour of the parameters 
describing the atmospheric turbulence at any site is crucial for the
 design of efficient systems. There are three basic  parameters relevant to AO 
design and operation:  Fried's parameter ({\it r}$_0$), the isoplanatic angle ($\theta_0$), and the coherence 
time ($\tau_0$). These parameters can be defined in terms of the refractive index structure constant profile 
($C_N^2(h)$) and the vertical wind profile ($V(h)$) (Roddier, Gilli \& Lund 1982):

\begin{equation}
r_0=\left[0.423k^2(\sec\zeta)\int{dhC_N^2(h)}\right]^{-3/5}
\label{r0}
\end{equation}

\begin{equation}
\theta_0=\left[2.914k^2(\sec\zeta)^{8/3}\int{dhC_N^2(h)h}^{5/3}\right]^{-3/5}
\label{AO0}
\end{equation}

\begin{equation}
\tau_0=0.314(\cos\zeta)\frac{r_0}{V_0}
\label{tau0}
\end{equation}

\noindent where $\zeta$ is the zenith angle, {\it k} is the optical wave number and $V_0$ is  the average velocity of the turbulence given by:

\begin{equation}
V_0=\left[\frac{\int{dh C_N^2(h) V(h)^{5/3}}}{\int{dh C_N^2(h)}}\right]^{3/5}
\label{v0}
\end{equation}

These parameters are convenient measurements of the strength, distribution and variation of the turbulence (see 
Hardy 1998 for a detailed introdution to adaptive optics for astronomical telescopes). 

Monitoring programs of turbulence structure at 
 astronomical sites are therefore mandatory for obtaining the input parameters for
the design and operation of efficient AO systems providing high Strehl ratios. Nevertheless, data should be obtained over
 decades to obtain sufficient statistical significance. In order to overcome the lack
 of long-term information on turbulence structure at  astronomical sites,
 winds at the 200 mbar pressure level ($V_{200}$ hereafter) were proposed
as a parameter for estimating the total turbulence at any particular site.
 This proposal is based on the hypothesis that the integrated $C_N^2$ profile is strongly related to the peak of the atmospheric 
 wind vertical profile, which used to be at around the altitude of the 200 mbar pressure level (Vernin 1986).  The $V_{200}$ proposal 
 as a parameter for site AO capabilities was supported
 by the similar seasonal trend of the seeing and $V_{200}$ at Mauna Kea and La
 Silla  (Vernin 1986), and the results found at Cerro Pach\'on and
 Paranal  (Sarazin \& Tokovinin 2002,S\&T02 hereafter), where $V_0$ was
 found proportional to $V_{200}$: $V_0$ = 0.4 $\times V_{200}$ (S\&T02).
In addition,  a good correlation---of the form
 $V_0$ = 0.56 $\times V_{200}$---was also found above San Pedro M\'artir
 (Mexico) using an atmospheric model to simulate a large dataset of $C_N^2$ profiles (Masciadri \& Egner 2006, M\&E06 hereafter).

 Such a linear connection between $V_0$ and $V_{200}$ at any site---an assumption that is currently under discussion and being tested---would simplify
 the calculation of the input parameters for AO design. Henceforward, the
 problem could be reduced to determining statistics for the existing world-wide 
 long-term high-altitude winds data in climatological databases. Indeed,
 $V_{200}$ statistics has been already used as a parameter for ranking
 astronomical sites for their suitability for AO (Ilyasov, Tillayev \& Ehgamberdiev 2000; Sarazin 2002; Carrasco \& Sarazin
 2003; Chueca et al.\ 2004; Carrasco et al.\ 2005; Garc\'{\i}a-Lorenzo et al.\
 2005; Bounhir, Benkhaldoun, \& Sarazin 2008). 

Despite  poor empirical results connecting seeing and $V_{200}$
 (Vernin 1986), the idea of a relation between image quality and
 high-altitude wind speed is increasingly widespread among those of the
 astronomical community interested in AO. 

Different meteorological processes are responsible for generating turbulence in the atmosphere.
 The turbulence in the first kilometre above the ground level (the boundary layer) is caused by local 
 factors (Lee, Stull \& Irvine 1984). These factors are buoyant convection processes, such as thermals 
 rising produced by surface solar heating, and mechanical processes, such as wind shear produced by the
  surface friction in the wind speed or by the lee waves formed by mountains or other geographic effects 
  (Stull 1988). In the free atmosphere, turbulence generators are connected to the synoptic scales conditions 
  (Erasmus 1986). The combination of both contributions will determine the quality of sites for astronomical
   observations. Site testing studies have reported a connection between ground layer winds and image quality 
   (Chonis, Claver \& Sebag 2009; Lombardi et al.\ 2007; Varela, Mu\~noz-Tu\~n\'on \& Gurtubai 2001;
    Mu\~noz-Tu\~n\'on, Varela \& Mahoney 1998; Erasmus 1986), obtaining in general better seeing 
    measurements for smaller wind speeds and at a prevailing wind direction. 

In this paper, we study the connection of high-altitude and ground based winds
 to the atmosperic AO input parameters for the Teide Observatory (Tenerife,
 Spain) using $C_N^2$ profiles from G-SCIDAR measurements and wind profiles
 from direct balloon measurements.

\section{THE DATA}

The Teide Observatory (hereafter OT) is located at an altitude of 2390 metres on the 
island of Tenerife (Canary Islands, Spain) at latitude $28^0 18^{'}$ N and longitude 
$16^030^{'}$ W. The OT was considered as a candidate site for the European Extremely Large 
Telescope (E-ELT). It is only $\sim160$ km distant from one of the most important E-ELT site 
candidates,  Roque de los Muchachos Observatory (hereafter ORM), on the island of La Palma 
(Canary Islands, Spain). The altitude of the ORM is also $\sim2400$ metres above sea level. 
The G-SCIDAR (Generalized
 SCIntillation Detection And Ranging) technique (e.g.\ Fuchs, Tallon
 \& Vernin, 1994) has been used to monitor the turbulence
 structure above OT and ORM (Garc\'{\i}a-Lorenzo, Fuensalida \&
 Rodr\'{\i}guez-Hern\'andez 2007) since November 2002,  both observatories showing a quite similar 
 seasonal behaviour in their average turbulence profiles (Garc\'{\i}a-Lorenzo et al.\ 2009). The 
 statistical predominance of synoptic scaled phenomena has been proposed to explain the observed 
 similarities of turbulence structure at both sites (Castro-Almaz\'an, Garc\'{\i}a-Lorenzo \& Fuensalida 2009). 

The current database of C$_N^2$ profiles above the OT includes useful data for
more than 150 nights. The velocities of the turbulence layers
can be obtained from G-SCIDAR data (Garc\'{\i}a-Lorenzo \& Fuensalida 2006;
 Prieur et al.\ 2004; Avila et al.\ 2003; Kluckers et al.\ 1998). Nevertheless,
the G-SCIDAR technique provides wind speed measurements only where a turbulence
 layer is detected and we can eventually miss information due to temporal
 decorrelation of the scintillation and/or fluctuations of velocities during
 the integration time of the G-SCIDAR exposures (Avila et al.\ 2001).

Alternatively, for this site wind vertical profiles can be obtained from a close radiosonde 
station placed 13 km  from the OT. In G\"uimar (Lat: 28.46N, Lon:16.37W)
 on the island of Tenerife (Spain), Spain's Agencia Estatal de Meteorolog\'{\i}a 
(AEmet) launches radiosondes and is one of
 the stations of the NOAA database (Station 60018). The radiosondes are
 launched from an altitude of 105 metres  above mean sea level and reach an
 altitude of about 30 km, maintaining a nearly steady rate of ascent ($\sim3$ ms$^{-1}$). The balloons provided twice-daily measurements (at
 00UT and 12UT) of meteorological variables (including wind measurements) above
 this location, although we have only used the data from midnight. A remarkable
 correspondence between the turbulence layer velocities derived from G-SCIDAR
 at the OT and balloon measurements from station 60018 has been already 
reported (Garc\'{\i}a-Lorenzo \& Fuensalida 2006).

\begin{figure}
\centering
\includegraphics[scale=0.20]{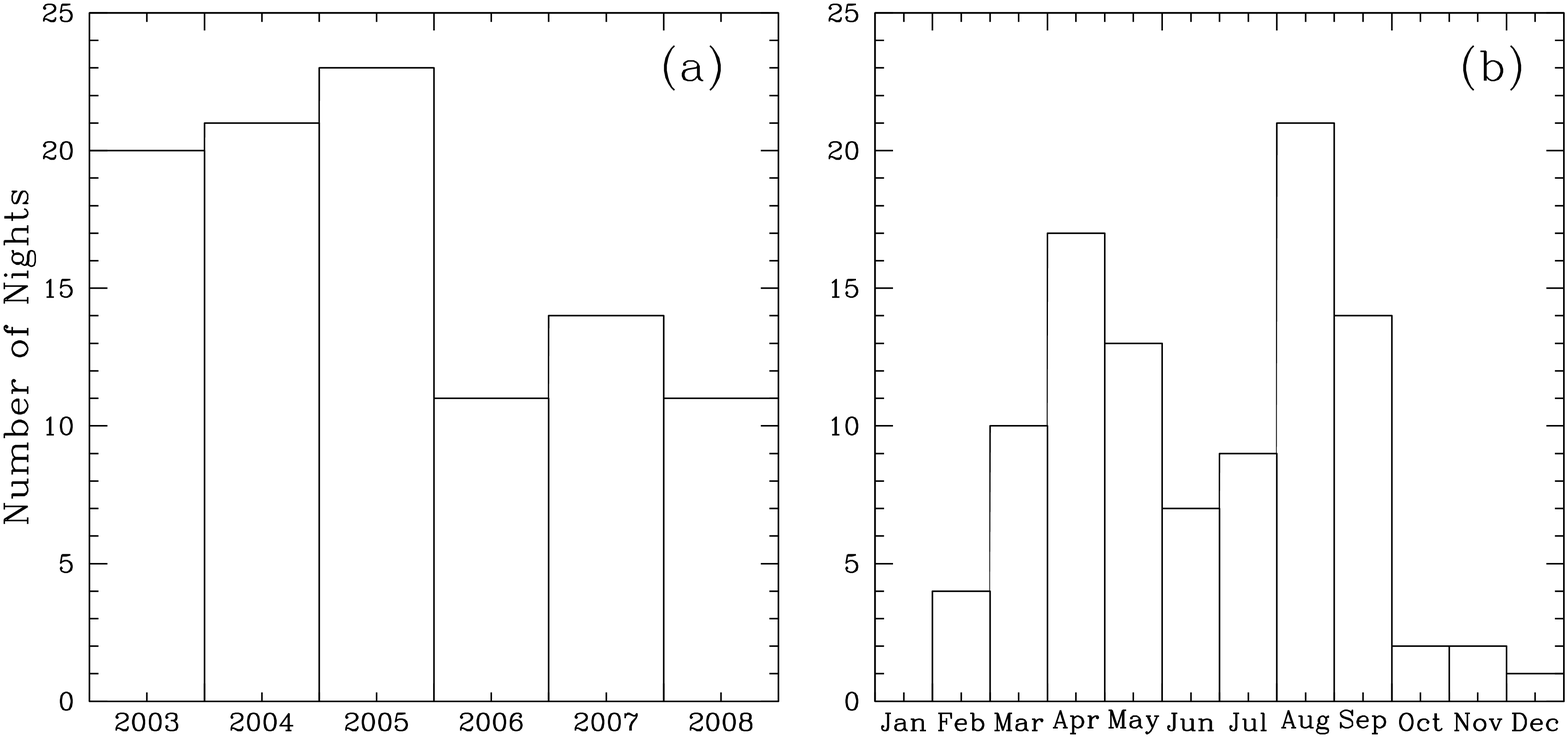}
 \caption{Distribution ((a) by year; (b) by month) of the nights used to study the relationship of wind speed and AO parameters at the OT.}
\label{dis_anos}
\end{figure}

We have used G-SCIDAR and balloon data for 100 nights to study relations between AO parameters 
and wind speeds at the OT. In order to calculate such parameters, we derived an average refractive index 
structure constant profile from  each 
 $CC_N^2(h)$ value obtained from 00UT to 02UT, approximately during the balloon ascent. The 100 nights included 
 in this study are distributed from 2003 to 2008 as is shown in Figure \ref{dis_anos}. The sample includes 
 more  nights in spring and summer than in autumn and winter. The reason for such a seasonal distribution
  has been imposed by the need to have simultaneous radiosonde and G-SCIDAR observations. Appendix A presents the 
  $C_N^2(h)$ and wind profiles for the 100 nights that constitute the data sample of this study. All the individual 
  $CN^2$ profiles habve been corrected for dome seeing using the method proposed by Fuensalida, Garc\'{\i}a-Lorenzo \& Hoegemann (2008).
Appendix B presents the dates of the different nights in the dataset and the total number of individual 
profiles---recorded simulateously with balloon data---used to obtain the average $C_N^2$ profiles. 

\begin{figure}
\centering
\includegraphics[scale=0.25]{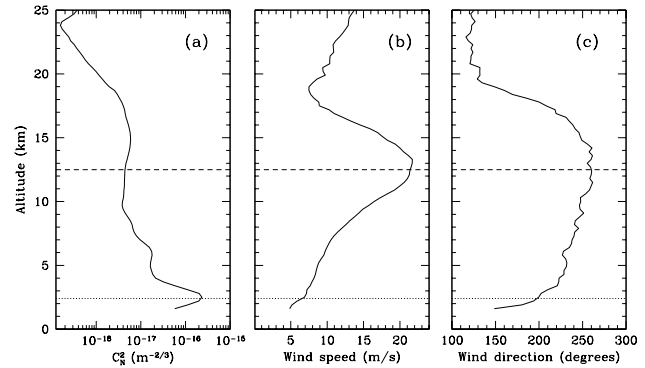}
 \caption{(a) Average of the $C_N^2$ measurements obtained between 00 UT and 02 UT at the Teide Observatory during 100 nights. 
 The number of individual profiles used to compute this average profile is greaater than 20200. (b) Statistical wind vertical 
 profile (average) derived from 100 balloon flights launched simultaneously with G-SCIDAR observations. Average wind at observatory 
 level ($\sim 2400$ m) could be not coincident with the average value derived from meteorological stations at the OT. 
 (c) Average wind direction derived from the balloon data for the 100 nights in the analysed dataset.  Average wind direction 
 at observatory level ($\sim 2400$ m) could be not coincident with the average value derived from meteorological stations at the OT. 
 The dotted line indicates the Teide Observatory level $\sim2400$ m, while the dashed line corresponds to the mean altitude of the 200 mbar
  pressure level derived from the data.}
% \caption{(a) Average (solid line) and median (dashed line) of the C$_N^2$ measurements obtained between 00-02 UT at the Teide observatory during 100 nights. The number of individual profiles used to compute this average profile is larger than 20000. (b) Statistical wind vertical profile derived from 100 balloon flights launched simulataneous to G-SCIDAR observations. (c) Average wind direction derived from the balloon data for the 100 nights in the analized dataset. The windangle is taken (as usual in atmospheric physics) clockwise from north. A wind blowing from the west has an angle of 270 degrees. }
\label{perfiles_promedio}
\end{figure}

A linear spline algorithm has been used to interpolate wind measurements from
 balloons to the same altitudes as the $C_N^2$ profiles. $V_{200}$ data have been also obtained from the same 
 radiosonde measurements. At the level of the OT, wind data measurements from
a local weather station next to the telescope where the G-SCIDAR is installed
have been used (hereafter $V_{\rm ground}$). The wind speed at the OT altitude 
provided by the radiosondes has been replaced by $V_{\rm ground}$ in the profiles for further
 calculations because wind measurements at this altitude can be strongly
 affected by orography. Figure \ref{perfiles_promedio} shows the mean turbulence and wind profiles derived by averaging 
 the total number of individual profiles in the dataset. The average $C_N^2(h)$ (derived from more than 20200 individual 
 profiles) shows that most of the turbulence is concentrated at the observatory level. It also reveals the presence of a 
 turbulence layer at around 8 km above mean sea level and other two turbulence features at $\sim8.5$ and 15 km 
 (Figure \ref{perfiles_promedio}a). This statistical profile is in good agreement with the turbulence structure derived 
 for a much larger database of $C_N^2$ profiles reported for the Canary Islands astronomical obsevatories (Garc\'{\i}a-Lorenzo et al.\ 2009 
 and references therein). Although not all the wind vertical profiles show a clear peak at 200 mbar pressure level (see individual profiles 
 in appendix A), the average $V(h)$ profile (Figure \ref{perfiles_promedio}b) shows its largest velocity at $\sim 13300$ m, 
 close to the mean altitude of the 200 mbar pressure level ($\sim 12500\pm 1200$ m). 
 This average wind vertical profile derived from the individual balloon measurement for the 100 nights in the dataset is 
 in agreement with the statistical $V(h)$ derived from a largest database (Garc\'{\i}a-Lorenzo et al.º 2005). 
 The obtained wind direction mean profile (Figure \ref{perfiles_promedio}c) shows a rotation in the wind direction from 
 dominant southern winds at low altitude to dominant western winds at 200 mbar pressure level and coming back to southerly prevailing
  winds at higher altitudes above the tropopause.

\section{DATA ANALYSIS}

The data available for the OT allow us to derive representative estimates of  Fried's parameter, 
the isoplanatic angle and the coherence time. In any case, the reader should take into account that we are 
using G-SCIDAR data from part of a night ($\leq$ than two hours) in order to match $C_N^2$ 
and radiosonde data during the balloon ascent, as we said in the previous section. Results could 
 differ slightly using data from  full nights. In order to calculate the integrals in equations 
\ref{r0}, \ref{AO0} and  \ref{v0} we used the well-known trapezoidal rule. Appendix B includes the 
individual values of the AO parameters (Fried's parameter,  theisoplanatic angle and the coherence time) at 
the OT for the hundred nights in the dataset. The average values for these parameters obtained including 
all measurements are $|r_0|$=15.30$\pm3.40$ cm, $|\theta_0|=2.84\pm1.13$ arcsec and $|\tau_0|=5.81\pm3.03$ ms,
 while the median values are $|r_0|$=14.81 cm, $|\theta_0|=2.56$ arcsec, and $|\tau_0|=4.97$ ms. 
 The uncertainties  indicate only the standard deviation of the averaged measurements. These statitical 
 values for AO parameters confirm the excellent sky quality of the Canary Islands astronomical sites 
 for AO implementations. 

At this stage, we have all the necessary data to study any possible connection between wind speed and AO parameters. 

\subsection{Connection between seeing and wind speed}

\begin{figure*}
\centering
\includegraphics[scale=0.42]{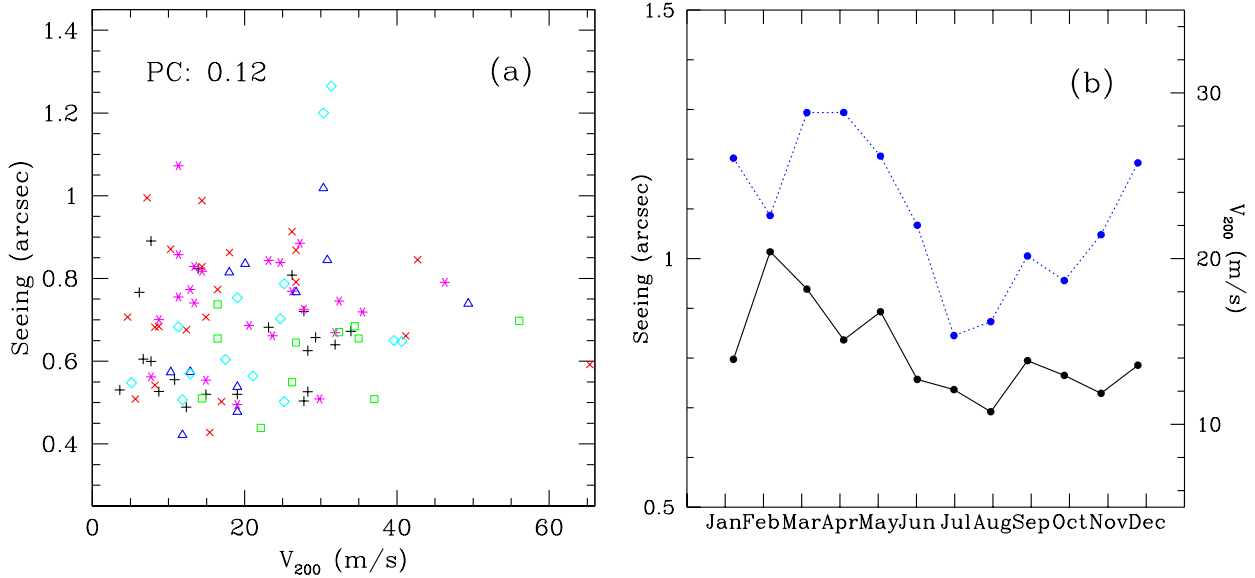}
 \caption{(a) Comparison of seeing and wind speed at 200 mbar pressure level at the OT. The linear Pearson 
 correlation coefficient between both quantities is shown at the top-left. Symbols indicate data from 
 different years: 2003--black crosses; 2004--red Xs; 2005--magenta asterisk; 2006--blue triangles; 2007--cyan 
 rhombus; and 2008--green squares. (b) Seasonal trend of the seeing (black line) and high-altitude winds ($V_{200}$, 
 blue-dashed line) above  Roque de los Muchachos Observatory ($\sim160$ km  from the OT and similar 
 altitude above mean sea level, $\sim$2400 m) for the period 1995--2002. }
\label{seeing_v}
\end{figure*}

 Vernin (1986) found a similar seasonal trend between seeing and $V_{200}$ at La Silla (Chile) 
 and Mauna Kea (Hawaii, USA), suggesting a possible connection between both variables.  Although 
 the $V_{200}$ and seeing connection has not been yet intensively checked at any site, the unfeasible 
 idea of identifying high-altitude wind speed with total seeing has become increasingly popular among 
 the astronomical community. In this section we study such a possible connection at the OT, deriving the 
 seeing from its relation to the Fried paramter: seeing = 0.98$\lambda$/$r_0$. Figure \ref{seeing_v}(a) presents 
 the comparison of the mean seeing derived from G-SCIDAR profiles for the 100 nights in our dataset 
 (see individual $r_0$ values in Appendix B) with the $V_{200}$ measurements (see Appendix B) obtained 
 from the radiosonde for the same nights and time lapses. This figure reveals a chaotic distribution of 
 data when comparing both variables. The Pearson correlation coefficient is close to zero, indicating 
 that a linear relationship between seeing and $V_{200}$ cannot be established at the OT.

However, a similar seasonal trend between $V_{200}$ and seeing could be still possible. Indeed, the 
seasonal behaviour of $V_{200}$ above the Canary Islands astronomical observatories (Garc\'{\i}a-Lorenzo
 et al.\ 2005; Chueca et al.\ 2004) reveals that the largest $V_{200}$ occurs in spring and the lowest in
  summer. This is consistent with the seeing behaviour reported for both the OT and the ORM sites, where the best
   seeing occurs in summer and the worst in spring (Mu\~noz-Tu\~n\'on, Vernin \& Varela 1997; Mu\~noz-Tu\~n\'on, 
   Varela \& Mahoney 1998). Although there are no seeing monitors regularly operative at the OT, a large 
   database of seeing measurements from DIMMs (Differential Image Motion Monitors) is available for the 
   ORM at the webpage of the Site Quality group of the Instituto de Astrof\'{\i}sica de Canarias 
   (http://www.iac.es/site-testing/). We have obtained all the seeing data for the ORM from 1995 to 2002 and we 
   have derived the seasonal evolution of seeing at this site (Figure \ref{seeing_v}(b)) by averaging all the 
   data obtained for each month. We have also derived the statistical $V_{200}$ behaviour throughout the year for 
   the same period (1995--2002) using a reduced sample of the $V_{200}$ timeserie analysed in 
   Garc\'{\i}a-Lorenzo et al.\ (2005) for ORM from climate diagnostic archive data. Figure \ref{seeing_v}b 
   show the seasonal evolution of $V_{200}$ in comparison with seeing behaviour for the period 1995--2002. Both 
   seeing and $V_{200}$ seem to follow a relative similar seasonal trend, as was found at Hawaii and La Silla (Vernin 1986). 

\begin{figure*}
\centering
\includegraphics[scale=0.4]{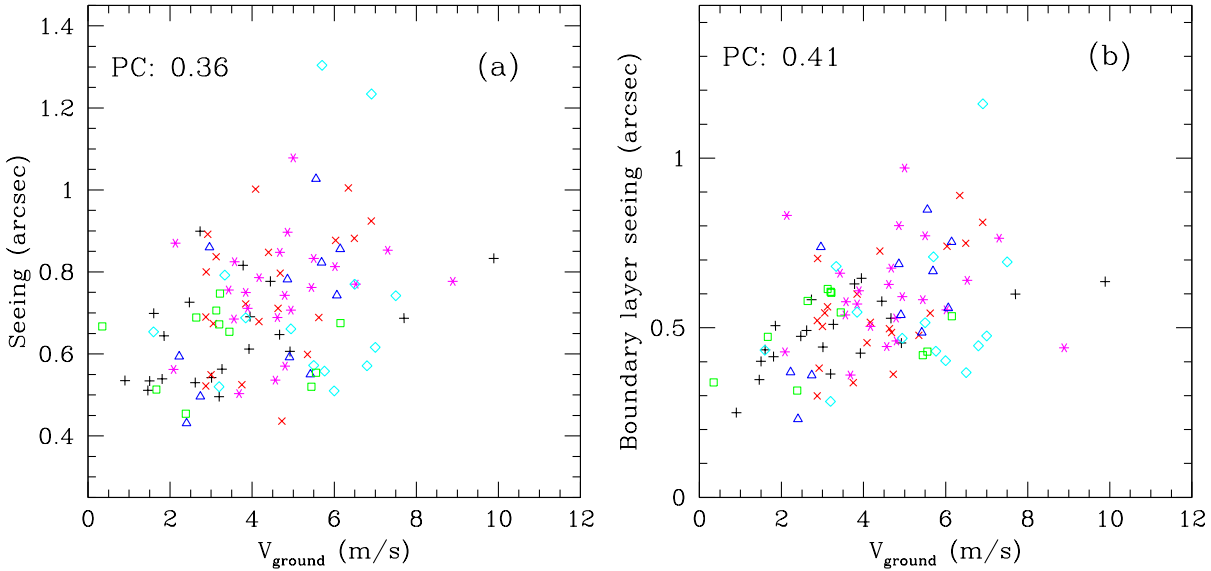}
 \caption{(a) Comparison of seeing and ground level wind speed for the OT.  The linear Pearson correlation 
 coefficient between both variables is shown at the top-left. Symbols indicate data from different years: 
 2003--black crosses; 2004--red Xs; 2005--magenta asterisk; 2006--blue triangles; 2007--cyan rhombus; and 2008--green 
 squares. (b) Comparison of boundary layer (first km) contribution to the seeing and wind speed at ground level 
 at the OT. The linear Pearson correlation coefficient between both quantities is shown at the top-left. Symbols 
 are as in (a). }
\label{seeing_vgb}
\end{figure*}

Previous studies have also suggested a possible connection between ground layer wind (speed and direction) to 
seeing (Chonis, Claver \& Sebag 2009; Lombardi et al.\ 2007; Varela, Mu\~noz-Tu\~n\'on \& Gurtubai 2001; 
Mu\~noz-Tu\~n\'on, Varela \& Mahoney 1998; Erasmus 1986). Such a relationship between seeing and $V_{\rm ground}$ 
seems to be present at OT (Fig. \ref{seeing_vgb}(a)) although a large dispersion in the data is clear. In general, 
larger seeing are obtained as $V_{ground}$ increases. The large dispersion of the data as well as the fact
 that the best fit derived (seeing=0.56+0.03V$_{ground}$) is far from the origin of coordinates could be
  related to the role played by wind direction. Such dependency of seeing on wind direction has been reported
   for El Pe\~n\'on  in Chile (Chonis, Claver \& Sebag 2009). Looking for an enhancement of this feature, M\&E06 
   studied the correlation between $V_{\rm ground}$ and the ground layer contribution to the seeing in San Pedro
    M\'artir. Figure 4(b) shows the comparision of wind speed at ground level to boundary layer (first km) 
    contribution to the seeing above the OT. A clear correlation between both quantities is present, showing 
    less dispersion than in the seeing--$V_{\rm ground}$ connection. In this case, the interception of the best linear 
    fit, boundary\_layer$_{\rm seeing}$ =0.39+0.04V$_{\rm ground}$, is still large, suggesting that another variable 
    (perhaps wind direction) is also playing an important role in the generation of turbulence at the boundary 
    layer.

\subsection{Relation between isoplanatic angle and wind speed}

\begin{figure*}
\centering
\includegraphics[scale=0.4]{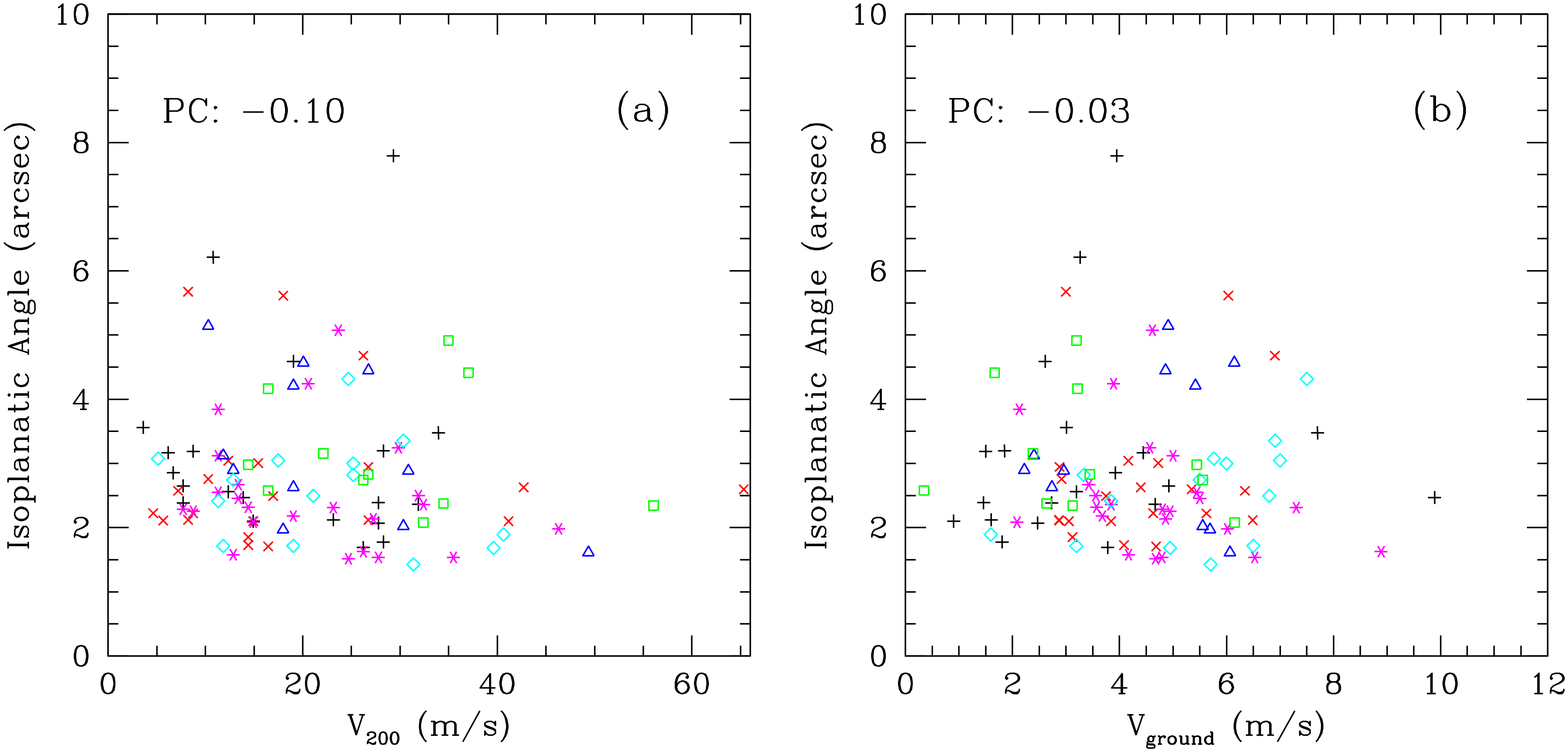}
 \caption{(a) Comparison of isoplanatic angles and 200 mbar pressure level wind speed for the OT. (b) Comparison of 
 isoplanatic angles and wind speed at ground level at the OT. The linear Pearson correlation coefficient between variables 
 is shown at the top-left of each pannel. Symbols indicates data from different years: 2003--black crosses; 2004--red Xs; 
 2005--magenta asterisk; 2006--blue triangles; 2007--cyan rhombus; and 2008--green squares. }
\label{AO_v}
\end{figure*}

A slow trend in the behaviour of the mean isoplanatic angle and $V_{200}$ was reported for Paranal during 2000 
(S\&T02), although a systematic relation between wind speed and $\theta_0$ seems to be non-existant. However,
 the astroclimatology webpage of the ESO sites (http://www.eso.org/gen-fac/pubs/astclim/paranal/seeing/adaptive-optics/) 
 suggests a connection between $V_{200}$ and isoplanatic angle showing the largest $\theta_0$ for the smallest $V_{200}$. 
 At the OT, we do not find any linear relation either with the wind speed at either the 200 mbar level or at  ground level and 
 isoplanatic angles obtained from G-SCIDAR profiles (Figure \ref{AO_v}). The Pearson correlation coefficients derived 
 comparing night-to-night $\theta_0$ with $V_{200}$ and $V_{\rm ground}$ are close to zero, indicating that there are no 
 systematic relations. Unfortunately, we do not have a  large enough database of isoplanatic angle measurements at the OT to 
 study any seasonal connection.
 
\subsection{Average velocity of the turbulence vs.\ wind speed at 200 mbar}

 $V_{200}$ was adopted as a parameter for site evaluation mainly thanks to the correlations found at Cerro Pach\'on 
 (S\&T02) and San Pedro M\'artir (M\&E06) between $V_0$ and $V_{200},$ as we already have mentioned. We explore 
 the $V_0$--$V_{200}$ connection in detail at the OT with our large dataset. Table \ref{best} presents the best linear 
 fit between $V_0$ and $V_{200}$ derived for each  year and including all the data.

%\begin{figure*}
%\centering
%\includegraphics[scale=0.65]{relacion_anos.eps}
% \caption{Comparison of the average velocity of the turbulence (V$_0$) to wind speed at 200 mbar pressure level (V$_{200}$) measurements for six consecutive years at the OT. Year is indicatedat the top-right of each plot. The linear Pearson correlation coefficient is shown at the top-left of each plot. Black continuos line corresponds to V$_0$=0.56$\times$V$_{200}$ approach found by M\&E06 for San Pedro M\'artir. Blue line indicate the best linear fit, while black dashed line is the fit forced to pass through the coordinate origin}
%\label{v0-v200}
%\end{figure*}

\begin{table}
\centering
\caption{Best linear fit obtained from the average velocity of the turbulence and wind speed at 
the 200 mbar pressure level measurements for the Teide Observatory. The best fit to the total sample 
(data from 2003+2004+2005+2006+2007+2008) is labelled  ``Total''. The last column corresponds to the 
result when forcing the fit to pass through the coodinate origin (``Best forced linear fit'')}
\begin{tabular}{|c|c|c|c|}\hline
{\bf Year} & {\bf Pearson's} & {\bf Best linear fit} & {\bf Best forced } \\
 &{\bf coefficient}  & & {\bf lineal fit } \\\hline
2003 &  0.23 &  7.68+0.07V$_{200}$ & 0.44V$_{200}$ \\
2004 &  0.74 &  6.35+0.16V$_{200}$ & 0.59V$_{200}$ \\
2005 &  0.58 &  5.34+0.18V$_{200}$ & 0.47V$_{200}$ \\
2006 &  0.77 &  3.86+0.23V$_{200}$ & 0.45V$_{200}$ \\
2007 &  0.70 &  5.25+0.27V$_{200}$ & 0.47V$_{200}$\\
2008 &  0.31 &  8.21+0.08V$_{200}$ & 0.41V$_{200}$\\\hline
{\bf Total} & 0.56 & 6.21+0.16V$_{200}$ & 0.47V$_{200}$\\\hline
\end{tabular}
\label{best}
\end{table}

The degree of correlation found at the OT (Table \ref{best}) indicates a faint linear relationship
 between $V_0$ and $V_{200}$ during 2003 and 2008, while Pearson's correlation coefficient for 
 measurements from 2004 to 2007 suggests a clearer linear connection between both quantities. The best 
 linear fit derived from yearly data are far from a linear fit passing through the coordinate origin, 
 as suggested for Paranal/Cerro Pach\'on (S\&T02) and San Pedro M\'artir (M\&E06). Indeed, we 
 found a relative large $V_{0}$ offset at the coordinate origin that ranges from 3.86 to 8.21 m/s 
 depending on year. In spite of this offset and following the same calculation as S\&T02 and M\&E06, 
 we have forced the fit to pass through the coordinate origin (Table \ref{best}). The proportional factor 
 between $V_{200}$ and $V_0$ ranges from 0.41 to 0.59 depending on year (see Table \ref{best}), giving
  a mean value of 0.47$\pm0.06$. This result indicates how such an approach to estimate $V_0$ could induce 
  large errors at the OT.

\begin{figure*}
\centering
\includegraphics[scale=0.4]{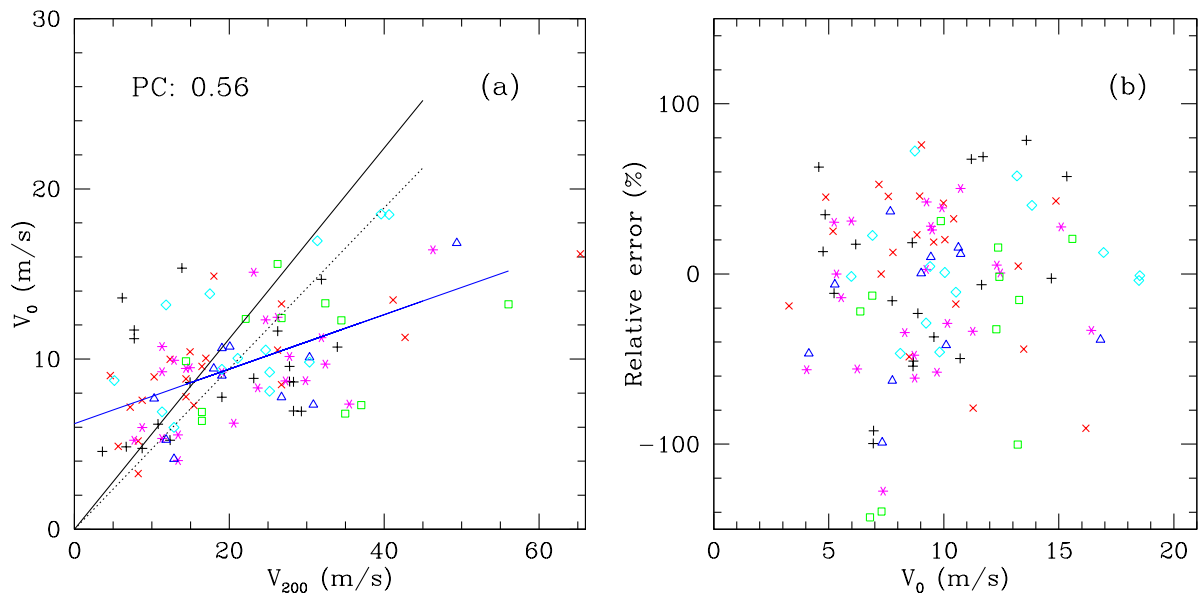}
 \caption{(a) Comparison of the average velocity of the turbulence ($V_0$) with wind 
 speed at the 200 mbar pressure level ($V_{200}$) measurements at the OT. The linear Pearson 
 correlation coefficient is shown at the top-left. The black continous line corresponds to 
 $V_0=0.56\times V_{200}$ approach found by M\&E06 for San Pedro M\'artir. The dashed black 
 line indicates the fit forced to pass through the coordinate origin, including the 100 measurements 
 from the sample. The best linear fit to the data ($V_0=6.21+0.16V_{200}$) is drawn as a blue line. 
 Symbols indicate data from different years: 2003--black crosses; 2004--red Xs; 2005--magenta asterisk; 
 2006--blue triangles; 2007--cyan rhombus; and 2008--green squares. (b) Relative error induced by using 
 the linear approach $V_0=0.47V_{200}$ at the OT for the full dataset. Negative values indicate 
 smaller estimations of $V_0$ than the measured. Symbols are as in (a).}
\label{v0-v200t}
\end{figure*}

Combining all the data in the sample (100 measurements), we obtain a Pearson correlation coefficient 
of 0.56, indicating the degree of linear correlation between V$_{200}$ and $V_0$ (Fig. \ref{v0-v200t}(a)).
 Again, the best linear fit to the total sample presents a large $V_0$ offset in the origin (Table \ref{best}). 
 The fit forced to pass through the coordinate origin gives 0.47 as the proportionality coefficient, 
 which is in agreement (assuming a dispersion of $\pm0.06$ in this parameter) with the 0.4$V_{200}$ and 
 0.56$V_{200}$ found for Cerro/Pach\'on (S\&T02) and San Pedro M\'artir (M\&E06), respectively.

Figure \ref{v0-v200t}(b) presents the relative errors, ($\frac{V_0-0.47V_{200}}{V_0}$)100\%, that we 
would obtain when estimating $V_0$ from $V_{200}$ using the linear factor including all the nights and 
forcing the fit to pass through the origin derived for OT. Such an approach would induce relative mean errors of $\sim$38\%, 
reaching uncertainties greater than 100\% in some cases.

\subsubsection{The turbulence characteristic altitude}

The  characteristic turbulence altitude gives the effective height of the dominant turbulence, increasing 
when the high-level turbulence dominates. The mean turbulence height can be calculated from:
\begin{equation}
\hat{H} = 0.314 \frac{r_0}{\theta_0}
\label{alturaequivalente}
\end{equation}
 r$_{0}$  and $\theta{_0}$ being the Fried parameter and the isoplanatic angle, respectively (Fried 1976). 
As both parameters can be derived from G-SCIDAR profiles, the characteristic turbulence  altitude can be calculated 
(see appendix B for daily measurements at OT). The mean $\hat{H}$ derived from the dataset is 3.87$\pm1.30$ km, 
suggesting that the turbulence at the OT is distributed in lower-altitude layers than in Paranal or Cerro Pach\'on, 
where the average turbulence characteristic altitude was found $\sim6.4$ km (S\&T02). 

\begin{figure*}
\centering
\includegraphics[scale=0.4]{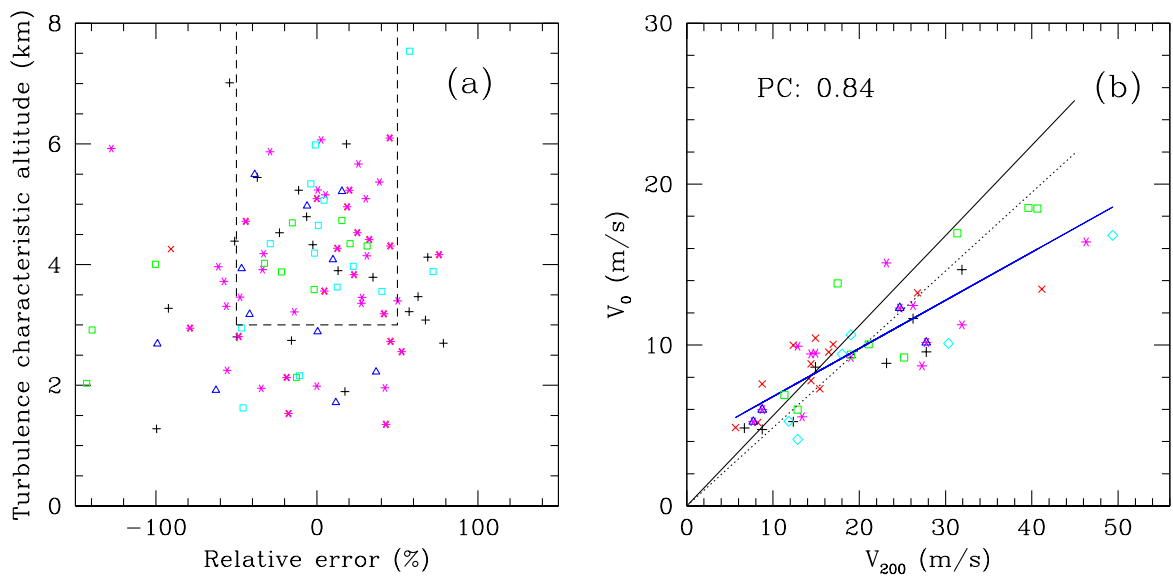}
 \caption{(a) Relative error (\%) (when assuming the expression $V_0=0.47\times V_{200}$ at the OT) versus the 
characteristic turbulence  altitude. The ashed line marks the region of nights satisfying the selection criteria 
 (see text). (b) Proportionality between $V_{200}$ and $V_0$ for selected nights at the OT. The best linear fit 
 to the data ($V_0=3.81+0.30V_{200}$) is drawn as a blue line. Symbols indicate data from different years: 
 2003--black crosses; 2004--red Xs; 2005--magenta asterisk; 2006--blue triangles; 2007--cyan rhombus; and 2008--green 
 squares. The linear Pearson correlation coefficient is shown at the top-left. The black continous line corresponds to 
 $V_0=0.56\times V_{200}$ approach found by M\&E06 for San Pedro M\'artir. The dashed black line indicates the fit 
 forced to pass through the coordinate origin  ($V_0=0.49V_{200}$) including the selected measurements. }
\label{crit2}
\end{figure*}

S\&T02 found that measurements providing errors larger than 50\% with respect to $V_0=0.4\times V_{200}$ correspond 
to those cases when the characteristic turbulence  altitude was smaller than 3 km. From the OT dataset and 
following the criteria suggested by S\&T02, we have selected those data with an equivalent turbulence height 
larger than 3 km and relative errors smaller than 50\% with respect to the $V_0$--$V_{200}$ proportionality derived for 
the OT ($V_0=0.47V_{200}$, Fig. \ref{crit2}(a)). Using these cut-offs, the number of selected nights is 56. The Pearson
correlation coefficient of $V_0$ and $V_{200}$ for this reduced dataset reaches 0.84, giving a proportionality factor of
 0.49 when forcing the fit to pass through the coordinate origin. The resulting $V_{200}$ vs.\ $V_0$ diagram 
 (Fig. \ref{crit2}(b)) shows less dispersion in the data, although the best fit of this sub-sample is still far 
 from passing through the origin (best fit: $V_0=3.81+0.30V_{200}$).

However, for 17\% of the nights in the total sample, the relative error respect $V_0=0.47V_{200}$ is larger 
than 50\% while the characteristic turbulence  altitude is larger than 3 km. Calculating the cumulative $C_N^2$ distribution, 
we found that for 15 of these nights,  70\% of the turbulence is concentrated under 4 km. For the remaining two  
nights, more than 50\% of the turbulence is in low-altitude layers (under 4 km). We have also noted that for  
14 of these 17 nights, the wind vector suffers significant twisters (larger than 50 degrees) from the ground to the 200-mbar 
level suggesting than wind direction could be influencing any connection.

Therefore, it seems that the linear relationship between $V_0$ and $V_{200}$ of the form $V_0=A\times V_{200}$ 
 can  be adopted only when the characteristic turbulence  altitude is larger than around 3 km, as indicated when 
comparing Figures \ref{v0-v200t} and Figure \ref{crit2}(b), and  when the wind direction  might be playing an important role 
to fix such an approach.

\subsection{Average velocity of the turbulence vs.\ wind speed at ground level}

\begin{figure*}
\centering
\includegraphics[scale=0.4]{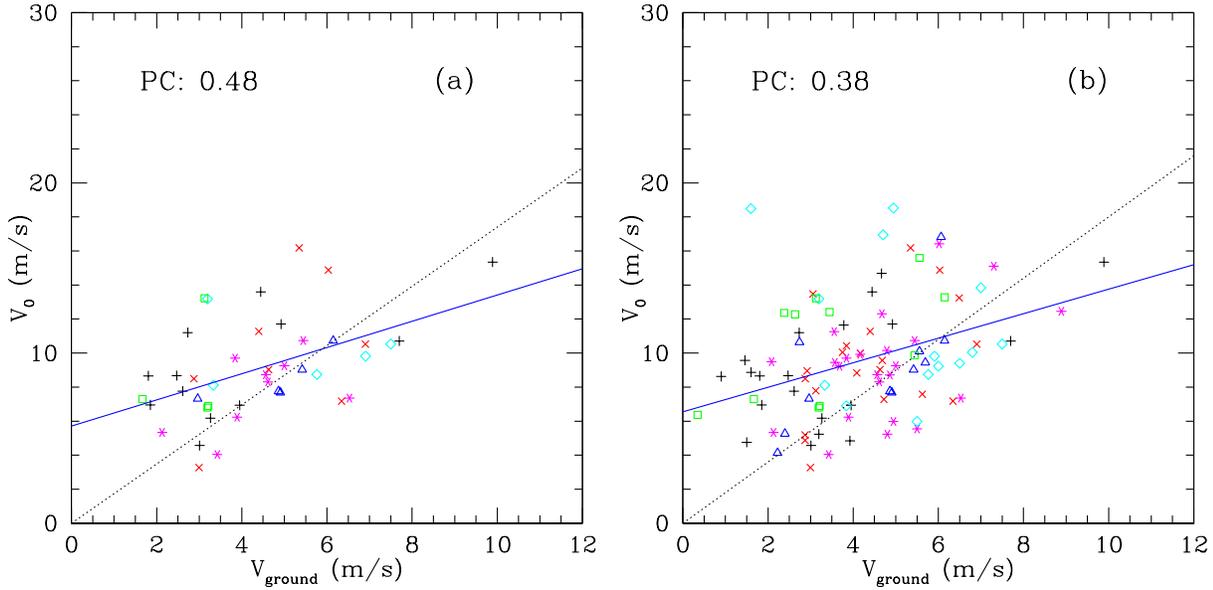}
 \caption{Comparison of the average velocity of the turbulence ($V_0$) with wind speed at  ground 
 level ($V_{\rm ground}$) measurements at the OT. The linear Pearson correlation coefficient is shown at 
 the top-left: (a) Selected data satisfiying that turbulence are mainly in low-altitude layers (see text). 
 The dashed black line indicates the fit forced to pass through the coordinate origin including the 44 measurements 
 of the sub-sample. The best linear fit to the data is drawn as a blue line; (right) Including all the data (100 
 nights).  The black continous line corresponds to $V_0=0.56\times V_{200}$ approach found by M\&E06 for San Pedro 
 M\'artir. The dashed black line indicates the fit forced to pass through the coordinate origin including the 100 
 measurements of the sample. The best linear fit to the data is drawn as a blue line ($V_0=7.89+0.45*V_{\rm ground}$). 
 Symbols indicates data from different years: (a) 2003: black crosses; (b) 2004: red Xs; (c) 2005: magenta asterisk; 
 (d) 2006: blue triangles; (e) 2007: cyan rhombus; and (f) 2008: green squares.}
\label{V0_Vground}
\end{figure*}

We have selected the data from the total dataset with relative errors larger than 50\% respect to the linear 
relationship $V_0=0.47V_{200}$, giving 44 nights. These data are characterized for characteristic turbulence  altitudes 
smaller than 3 km and/or a $C_N^2$ distribution mainly concentrated under 4 km. We have compared the $V_0$ measurement 
from this subset with the mean wind velocity at  ground level during G-SCIDAR observations. $V_{\rm ground}$ data are 
measured by a weather station placed closer to the Carlos S\'anchez Telescope where the G-SCIDAR data are collected. 
Figure \ref{V0_Vground}(a) shows $V_0$ as a function of $V_{\rm ground}$. The Pearson correlation coefficient between
 $V_0$ and $V_{\rm ground}$ is 0.48,  the best linear fit being $V_0=0.77V_{200}+5.71$. As in section \S3.1, we have 
 forced the fit to pass through the coordinate origin, giving $V_0=1.74V_{\rm ground}$. Such an approximation would induce 
 mean relative errors (abs($\frac{V_0-1.74V_{\rm ground}}{V_0}$)100\%) of 29\%, with a maximum error smaller than 65\%. 
 Considering all the nights in the dataset, the slope of the fit passing through the coordinate origin will be 1.80, 
 providing a mean relative error of 41\% and a maximum relative error smaller than 100\% (see Figure \ref{V0_Vground}(b)).
  This fact and the results in section \S3.2 suggest that $V_{200}$ and $V_{\rm ground}$ could provide estimates of
   $V_0$ with similar (or even bettter) uncertainties. 

S\&T02 proposed a general formula to estimate $V_0$ combining data from ground level and wind speed at 200 mbar pressure 
level: $V_0\simeq$ Max($V_{\rm ground}, 0.4V_{200}$). According to results in this section and \S3.2, such a connection should 
be of the form $V_0\simeq$ Max(1.74$V_{\rm ground}, 0.49V_{200}$) for the OT. This approach to estimating $V_0$ could also induce 
large errors at the OT, with a mean error of 28\%. Large uncertainties and the large intercepts derived for the best fits 
suggest that wind direction could be playing an important role in these connections.

Unfortunately, any of the relations to estimate $V_0$ provide in many cases large errors that are incompatible with the 
requirements for AO systems. Therefore, the on-site measurement of turbulence and wind profiles is still crucial for 
optimizing future AO instruments.

\subsection{Relation between coherence time and wind speed}

\begin{figure*}
\centering
\includegraphics[scale=0.4]{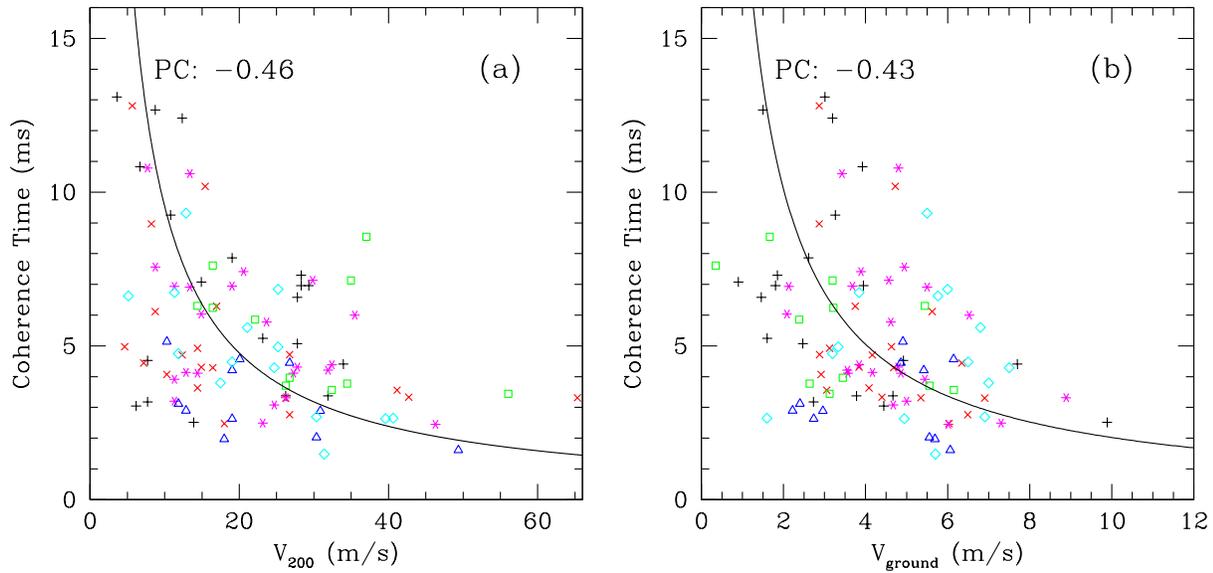}
 \caption{(a) Comparison of the coherence time of the turbulence ($\tau_0$) with wind speed at the 200 mbar presseure level ($V_{200}$) for 
 the OT.  The black continous line corresponds to the best linear fit: $\tau_0=8.35-0.12\times V_{200}$. (b) Comparison of $\tau_0$ 
 and wind speed at ground level. The linear Pearson correlation coefficient is shown at the top-left of each pannel. Symbols indicate 
 data from different years: 2003--black crosses; 2004--red Xs; 2005--magenta asterisk; 2006--blue triangles; 2007--cyan rhombus; and 2008--green 
 squares.}
\label{tau}
\end{figure*}

A connection between coherence time and wind speed is expected, given that a certain degree of relation exists 
beween $V_0$ and $V_{200}$ or $V_{\rm ground}$. Such a relation is clearly present at San Pedro M\'artir (M\&E06) and the
Chilean sites (S\&T02). Figure \ref{tau} shows the nightly mean values of $\tau_0$ in comparison to wind speeds at the OT. 
The derived Pearson correlation coefficients indicate a clear inverse connection between $\tau_0$ and wind speeds. As a general 
trend, the larger the wind speed is at ground level, the smaller coherence time obtained. We have fit a curve of the 
form $\tau_0=B$/wind, obtaining $B=20.21$ and $B=95.40$ for $V_{\rm ground}$ and $V_{200}$, respectively. 

Previous studies (S\&T02; M\&E06) actually proposed to estimate $\tau_0$ using the proportionality relation $V_0=A\times V_{200}$ 
in Equation \ref{v0}. We have estimated $\tau_0$ through the general approach derived in section \S3.4 to estimate $V_0$.  In this case, 
mean relative errors smaller than 15\% can be obtained, reaching uncertainties smaller than 50\% in all the nights in the dataset. 
According to this result, calculate $\tau_0$ from on-site seeing measurements and winds at ground and 200 mbar levels (through the 
general approach) provide relative good estimates of this AO parameter with acceptable uncertainties. 

\section{Discussion}

The wind speed at the 200 mbar pressure level has been accepted as an astronomical site evaluation parameter 
indicative of suitabiblity for adaptive optics. The hypothesis to propose high-altitude winds as a 
parameter for site evaluation was that the integrated refractive index structure constant is strongly 
related to the maximum wind speed in the atmosphere that is reached near the 200 mbar pressure 
level (Vernin 1986). Such a hypothesis seems not to be true at the OT for night-to-night measurements, although 
a similar seasonal trend between seeing and $V_{200}$ exists as in Mauna Kea and La Silla. Such a hypothesis 
is not true at the OT in the sense that there are many nights in which the maximum wind speed in the wind vertical 
profile is far from the 200 mbar presseure level, or even that there does not exist a clear maximum in the profile (see wind
 vertical profiles in Appendix B). Only 39\% of the profiles present a peak at around the 200 mbar pressure 
 level ($\approx1500$ m above or below the altitude of the 200 mbar level). Wind speed at ground level 
 seems to have more impact on total and boundary layer seeing than $V_{200}$: the larger V$_{\rm ground}$ is 
 the larger are the seeing values that are expected, although a large dispersion in the measurements is present. Orography 
 and wind direction can have an important influence in increasing such dispersion. 

$V_{200}$ as a site evaluation parameter was supported by the empirical results at Cerro Pach\'on/Paranal 
(S\&T02) and San Pedro M\'artir (M\&E06), where $V_0$ was found to be proportional to $V_{200}$ in the form:  
$V_0 = A\times V_{200}$,  A being a constant ($A=0.4$ for Cerro Pach\'on/Paranal; and $A=0.56$ for San Pedro 
M\'artir). Such a relationship is very attractive as it will simplify the problem of knowing the relevant 
input numbers for AO that could be parameterized in term of $V_{200}$. However, at the Teide Observatory 
such a relationship seems not to be as smooth as at the Chilean or Mexican sites. Only when the mean altitude 
of the turbulence is larger than 3 km, $V_{200}$ seems to connect better to $V_0$, but even in this case 
wind direction could have an important influence. We found similar uncertainties when estimating 
V$_0$ from $V_{200}$ or $V_{\rm ground}$, suggesting than both wind speeds can be used to simplify the calculation 
of $V_0$ if we are able to assume errors larger than 50\% in many cases. In any case, if a gross estimate 
of $V_0$ is required, the use of both wind speeds ($V_{200}$ and $V_{\rm ground}$) can provide better results. 

Including the wind speed, other factors (e.g.\ buoyant convection precesses, inestability phenomena, etc.) 
may be playing an important role in generating low/medium-altitude turbulence (Stull 1988) that could be 
breaking the connection with $V_{200}$. Moreover, changes in wind direction or wind regimes (in different 
seasons, for example) could have an important influence on the linear coefficient connecting $V_{0}$ or 
seeing with $V_{200}$. Indeed, many of the nights in the sample with large discrepancies with respect to a linear 
behaviour with V$_{200}$ show  significant wind direction gradients from ground to high-altitude levels (see 
wind direction profiles in Appendix B). The mean wind vector twist from low- to high-altitude levels at the 
OT (see section \S2) and the result could be a break on the influence of high-altitude winds, concentrating 
the turbulence generators at lower altitudes than in Paranal/Cerro Pach\'on. Such a twist in wind directions 
is smoother at the Chilean sites but is more important for the Hawaiian islands (Eff-Darwich et al.\ 2009). If wind 
direction changes are really playing an important role, we would expect the faintest night-to-night connection
 between $V_0$ and $V_{200}$ at the Mauna Kea site.

Such a break in the connection between AO parameters and high-altitude winds could be also related with the 
large variation of the tropopause level above the Canary Islands astronomical sites, ranging from $\sim200$ to 
100 mbar depending on season (Garc\'{\i}a-Lorenzo, Fuensalida \& Eff-Darwich 2004). Indeed, during winter and 
spring the tropopause level could be at a lower-altitude than the 200 mbar pressure level. This means that, 
depending on season, the 200 mbar pressure level could be in the stratosphere instead of the troposphere at 
Roque de los Muchachos or Teide Observatories. In contrast, the tropopause level above Paranal or Mauna 
Kea is always at higher altitudes than the 150 mbar pressure level (Garc\'{\i}a-Lorenzo, Fuensalida \& Eff-Darwich 2004). 

  Despite  poor empirical results, the false idea of a connection between image quality and high-altitude wind 
  speed has become increasingly widespread among those in the astronomical community interested in AO. Unfortunately, 
  all of the  connections found between winds and AO parameters are relatively faint, and their relations could 
  induce large errors in many cases that are not compatible with the requirements for efficient AO systems. 
  We would like to emphasize the importance of atmospheric on-site turbulence information to evaluate the 
  capabilities of adaptive optics and multi-conjugate adaptive optics systems. Any average or approached 
  value is only a gross estimate of the AO input parameters. If efficient AO is required, the correction 
  system should be prepared for a large variety of  turbulence conditions that might be very different from 
  night to night. Therefore, the measurement of turbulence and wind profiles are still crucial for optimizing 
  future instruments with extreme AO capabilities.

In any case, we have demostrated that any factor derived at a site to simplify calculations cannot be easily 
generalized worldwide but  should be obtained for each site. Indeed, the connections between parameters at a 
particular site might not be valid for any other site due to the pecularities of each location (latitude, 
longitude, orography, etc.). Any astronomical site evaluator should be checked carefully before any approach is adopted.

\section{Summary and Conclusions}

We have studied the connection between AO atmospheric parameters and wind speed at the ground and 200 
mbar pressure levels by means of a database of 100 nights of G-SCIDAR $C_N^2$ profiles and balloon data at 
the Teide Observatory. We have obtained the average profiles ($C_N^2(h)$ and $V(h)$) above this site and we 
derived the average seeing, isoplanatic angle and coherence time that confirm the excellent conditions at the OT 
for AO applications. Our main conclusions can be summarized as follows.

\begin{itemize}
\item[1.] The connection between night-to-night seeing and wind speed at the 200 mbar pressure level is very faint 
or non-existent at the OT. However, we found a similar seasonal trend between statistical seeing behaviour and $V_{200}$
that may suggest that $V_{200}$ only plays a secondary role.

\item[2.] The night-to-night seeing seems to be connected to ground level winds, although a large dispersion 
is obtained which may be related to the influence of other factors (wind direction, buoyant convection processes, etc.). 

\item[3.] A correlation between wind speed at the site level and the boundary layer contribution to seeing is 
found. Again, the wind direction could be a factor playing an important role in such a connection.

\item[4.] The isoplanatic angle is not connected to wind speeds, although a seasonal trend between both 
variables cannot be roled out.

\item[5.] The linear connection between the average velocity of the turbulence and wind speed at the 200 mbar 
pressure level is faint at the OT. Only in those cases where the average altitude of the turbulence is larger 
than 3 km can a better connection  be accepted.

\item[6.] Similar uncertainties are obtained when estimating $V_0$ from winds at ground level or from $V_{200}$. 
In both cases, the wind direction could play an important role.

\item[7.] The coherence time presents an inverse linear connection to winds, although it shows a large dispersion.

\item[8.] Best results are obtained when combining wind speed at ground and the 200 mbar pressure level. 

\item[9.] The large errors derived from any relation between AO parameters and winds are not compatible with 
the requirements of efficient AO systems. Moreover, such errors could be also a problem when $V_{200}$ is used 
as a site evaluator.
\item[10.] The proper characterization of atmospheric turbulence and winds is still crucial for optimizing 
future instruments with extreme AO capabilities.
\end{itemize}

The results in this paper indicate that adaptive optics parameters present complex connections to wind speed.

\section*{Acknowledgments}

The authors thank T. Mahoney for his assistance in editing this paper.
This paper is based on observations obtained at the Carlos S\'anchez Telescope operated by the Instituto de Astrof\'{\i}sica de Canarias at the Teide
 Observatory on the island of Tenerife (Spain). The authors thank all the staff at the Observatory for their 
 kind support. Radiosonde data were also used and recorded from the webpage of the Department of Atmospheric 
 Science of the University of Wyoming (http://weather.uwyo.edu/upperair/sounding.html). Balloons are launched 
 by the Spanish ``Agencia Estatal de Meteorolog\'{\i}a''. Seeing data from the Roque de los Muchachos Observatory 
 provided by the Instituto de Astrof\'{\i}sica de Canarias at the web page http://www.iac.es/site-testing/ have
   also been used. This work has also made use of the NCEP Reanalysis data provided by the National Oceanic and
   Atmospheric Administration-Cooperative Institute for Reasearch in Environmental Sciences (NOAA-CIRES) Climate 
   Diagnostics Center, Boulder, Colorado, USA, from their web site at http://www.ede.noaa.gov.

 This work was partially funded by the Instituto de Astrof\'{\i}sica de Canarias and by the Spanish Ministerio de 
 Educacion y Ciencia (AYA2006-13682). B. Garc\'{\i}a-Lorenzo and A. Eff-Darwich also thank the support from the Ram\'on y 
 Cajal program by the Spanish Ministerio de Educacion y Ciencia.

\newpage

\section{APPENDIX A}

This appendix includes the average C$_N^2$(h) profiles derived from G-SCIDAR measurement from 00UT to 02UT for the 100 nights
 considered in this work. We also plot the wind speed vertical profile (module and direction) provide by balloons launched from 
an altitude of 105 meters at about $\sim13$ km from OT at 00UT. 

\begin{figure*}
\centering
\includegraphics[scale=0.70]{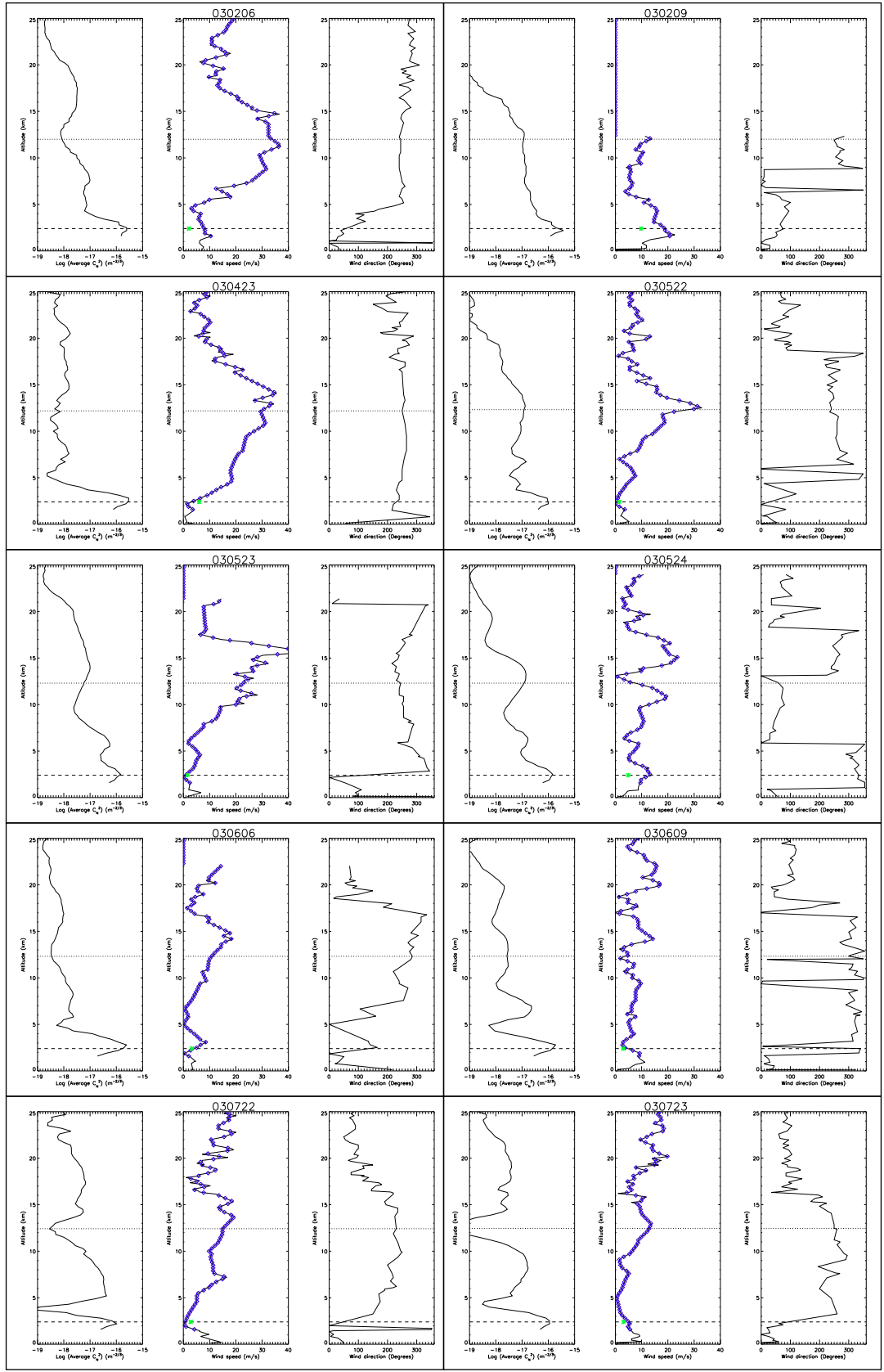}
 \caption{Average C$_N^2$ profile and the simultaneous wind vertical profiles (modulus and direction) for the data corresponding to 2003-2008 period at the Teide Observatory. Dates are indicated at the top of each plot. Open squares are the interpolated velocities to the same resolution than turbulence profiles. }
\label{todos_perfiles}
\end{figure*}

\begin{figure*}
\centering
\addtocounter{figure}{-1} 
\includegraphics[scale=0.70]{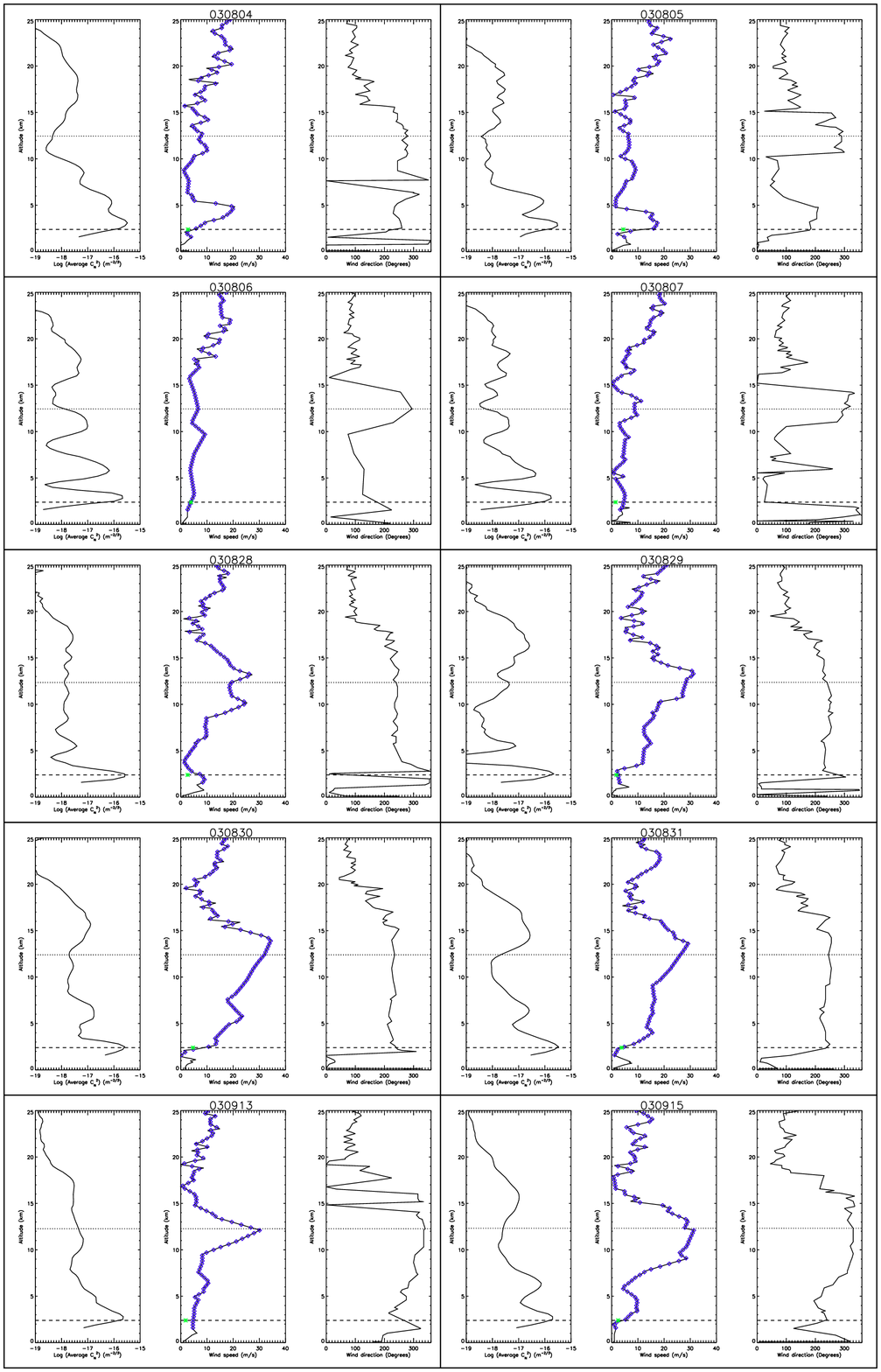}
 \caption{Continuation.  }
\label{todos_perfiles}
\end{figure*}

\begin{figure*}
\centering
\includegraphics[scale=0.70]{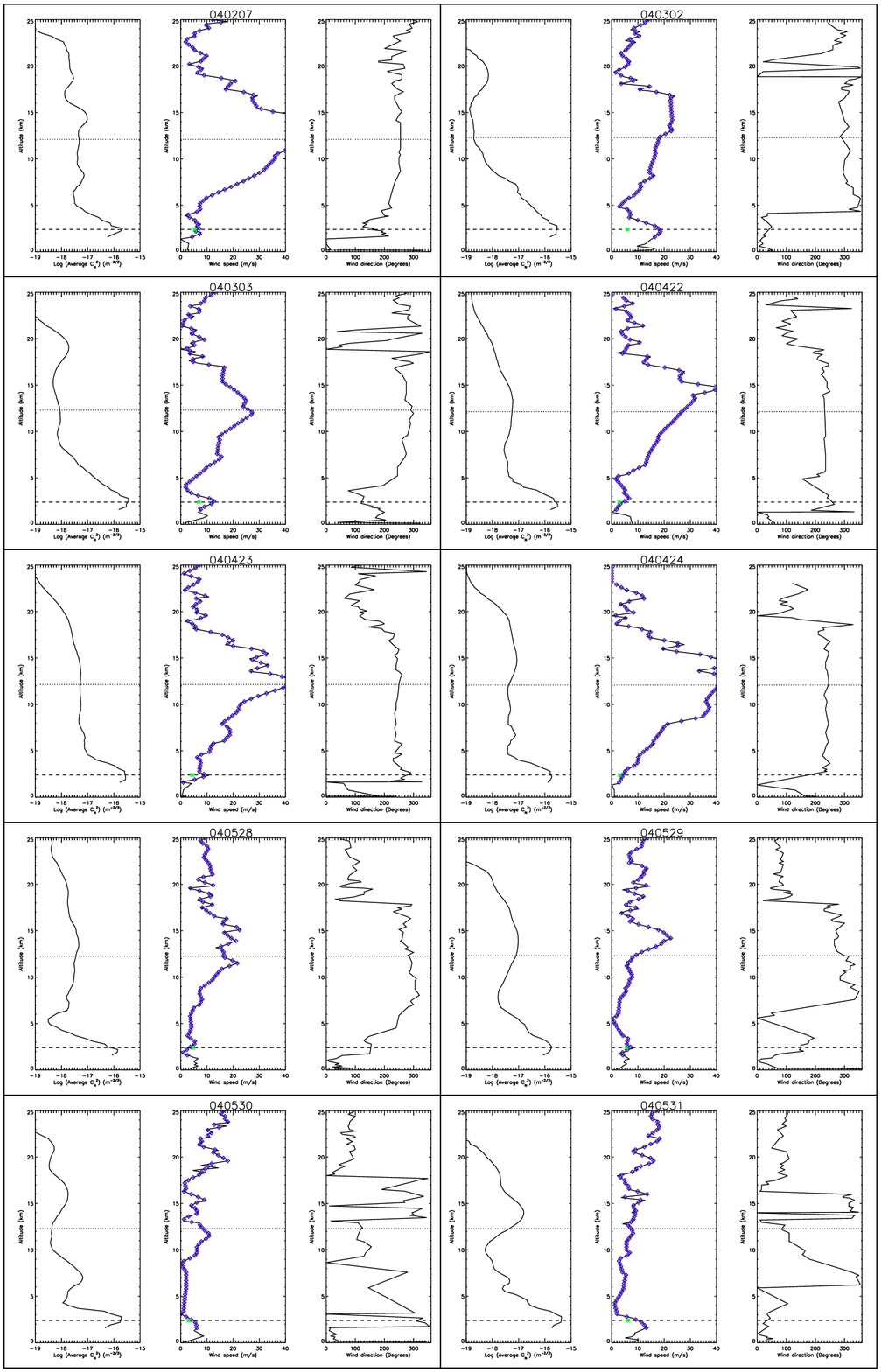}
 \caption{Continuation. }
\label{todos_perfiles}
\end{figure*}

\begin{figure*}
\centering
\addtocounter{figure}{-1} 
\includegraphics[scale=0.70]{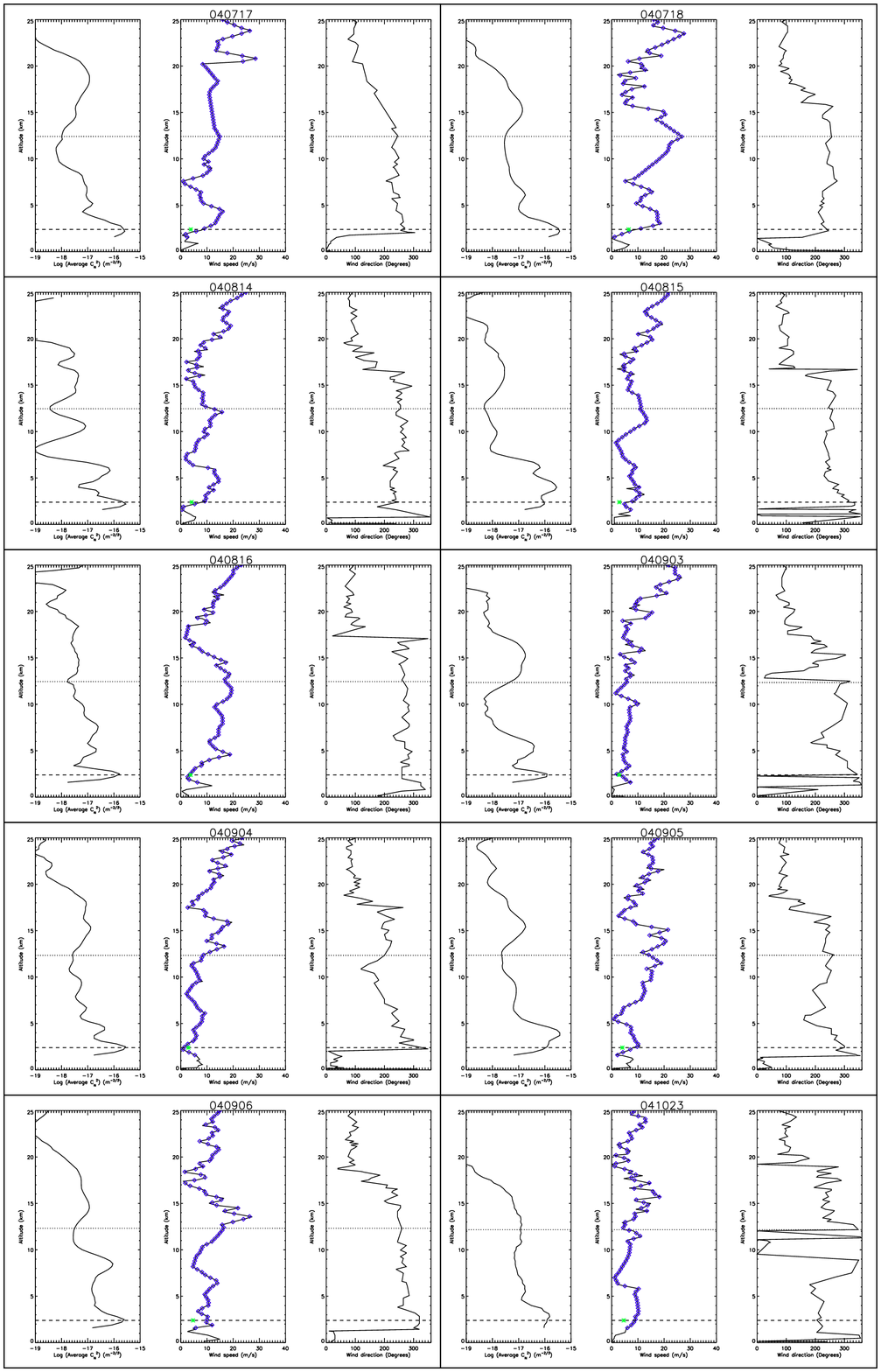}
 \caption{Continuation.  }
\label{todos_perfilesp}
\end{figure*}

\begin{figure*}
\centering
\includegraphics[scale=0.70]{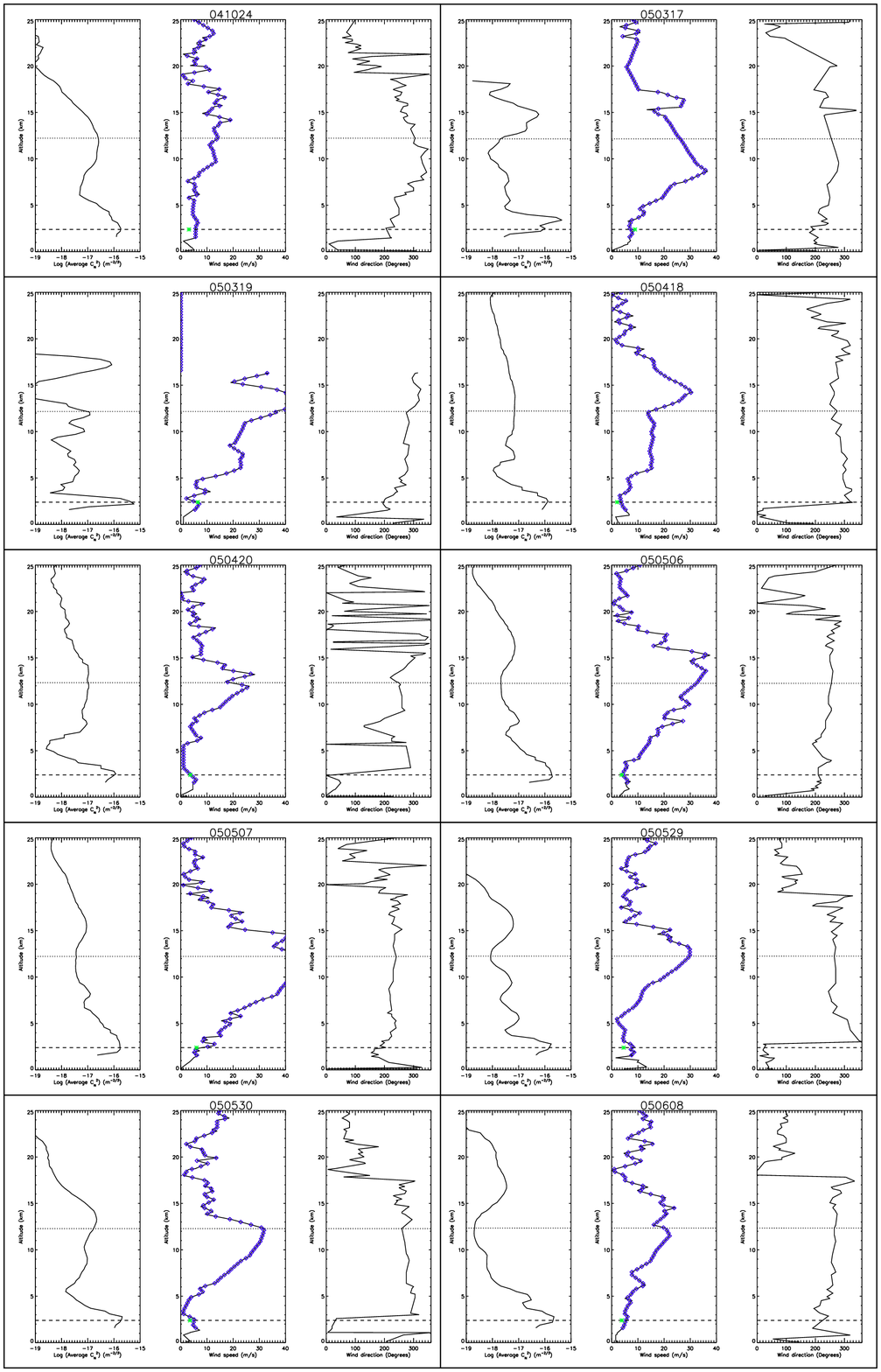}
 \caption{Continuation. }
\label{todos_perfiles}
\end{figure*}

\begin{figure*}
\centering
\addtocounter{figure}{-1} 
\includegraphics[scale=0.70]{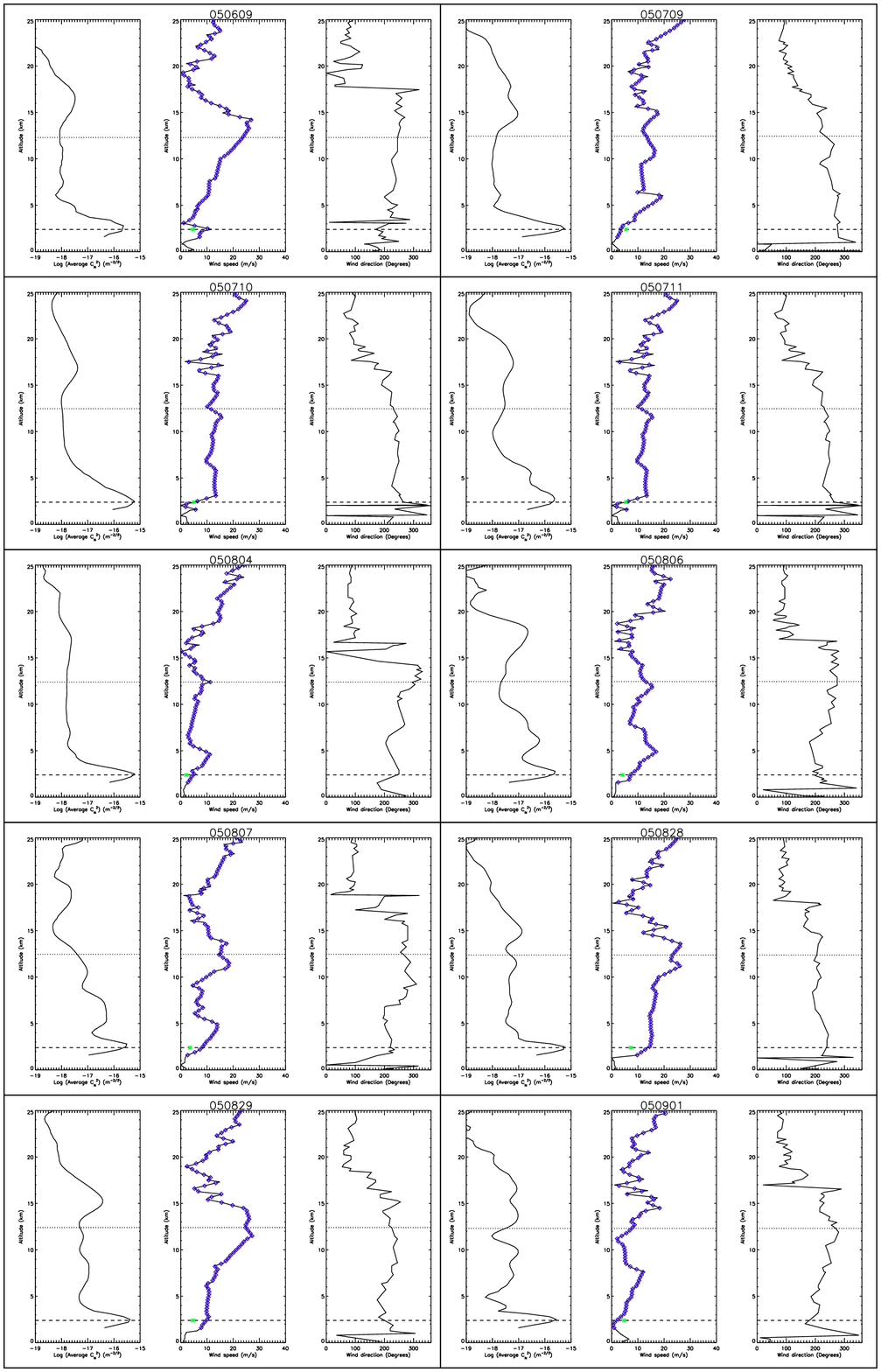}
 \caption{Continuation.  }
\label{todos_perfiles}
\end{figure*}

\clearpage

\begin{figure*}
\centering
\addtocounter{figure}{-1} 
\includegraphics[scale=0.70]{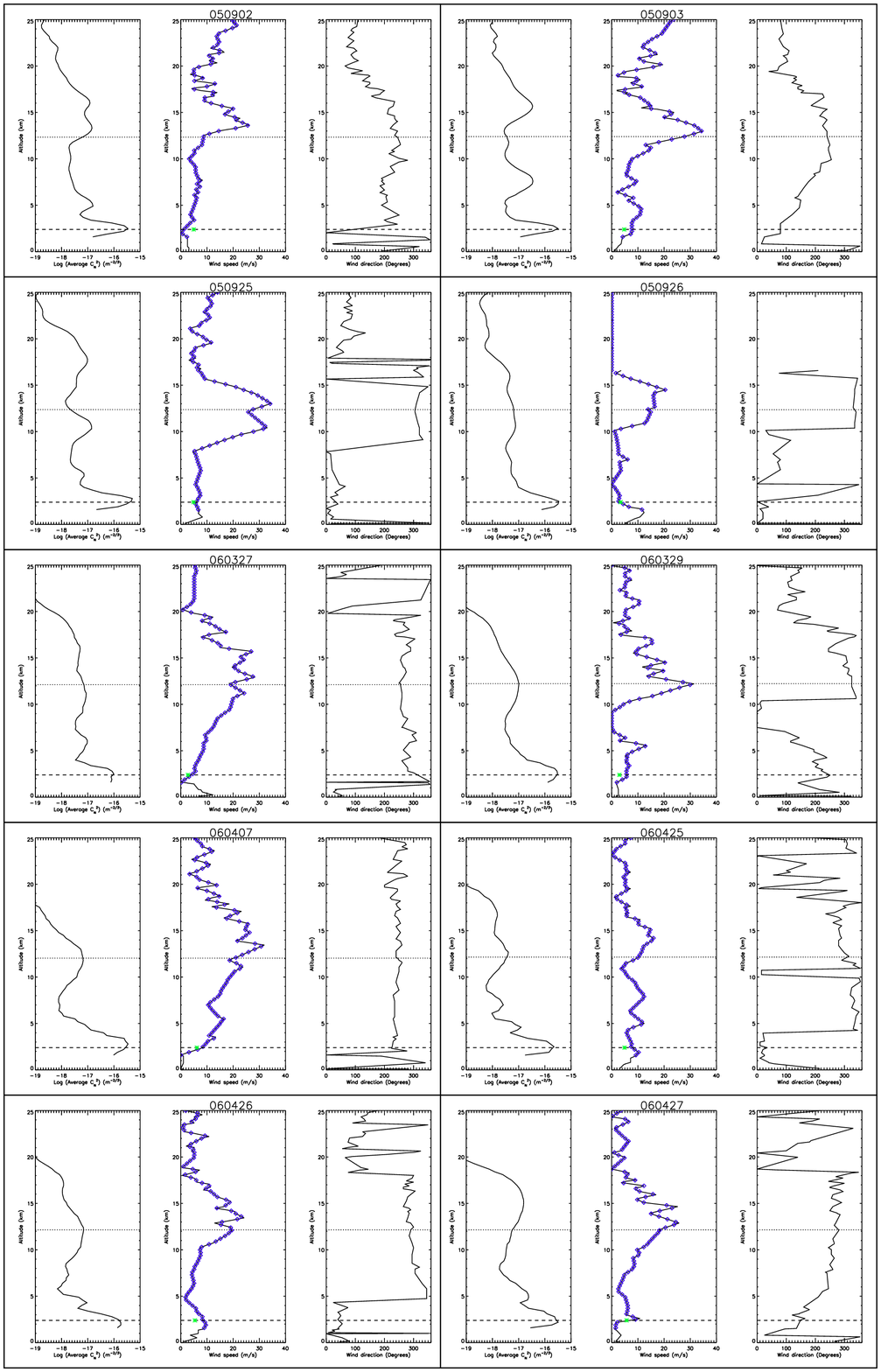}
 \caption{Continuation. }
\label{todos_perfiles}
\end{figure*}

\begin{figure*}
\centering
\addtocounter{figure}{-1} 
\includegraphics[scale=0.70]{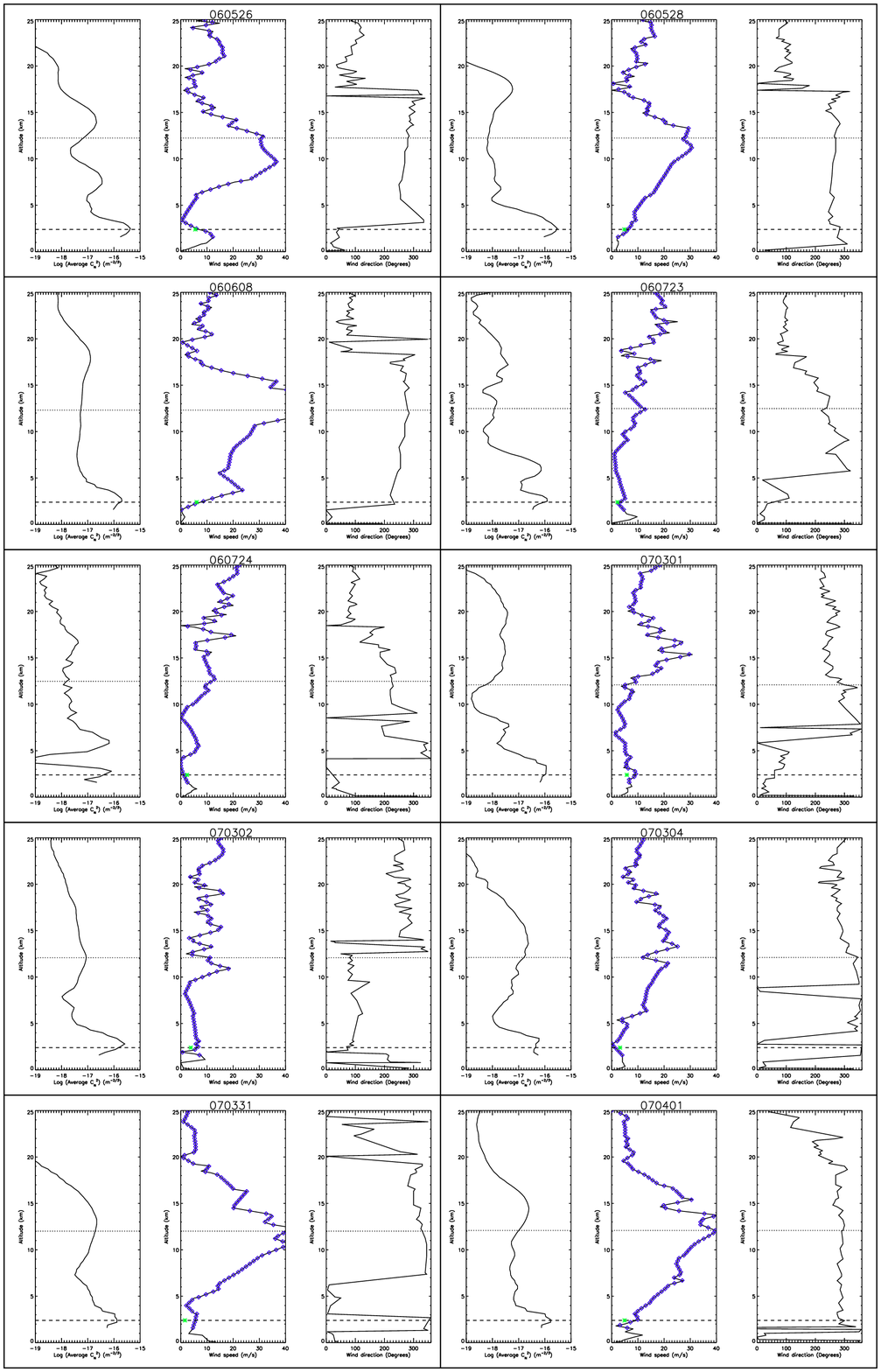}
 \caption{Continuation. }
\label{todos_perfiles}
\end{figure*}

\begin{figure*}
\centering
\addtocounter{figure}{-1} 
\includegraphics[scale=0.70]{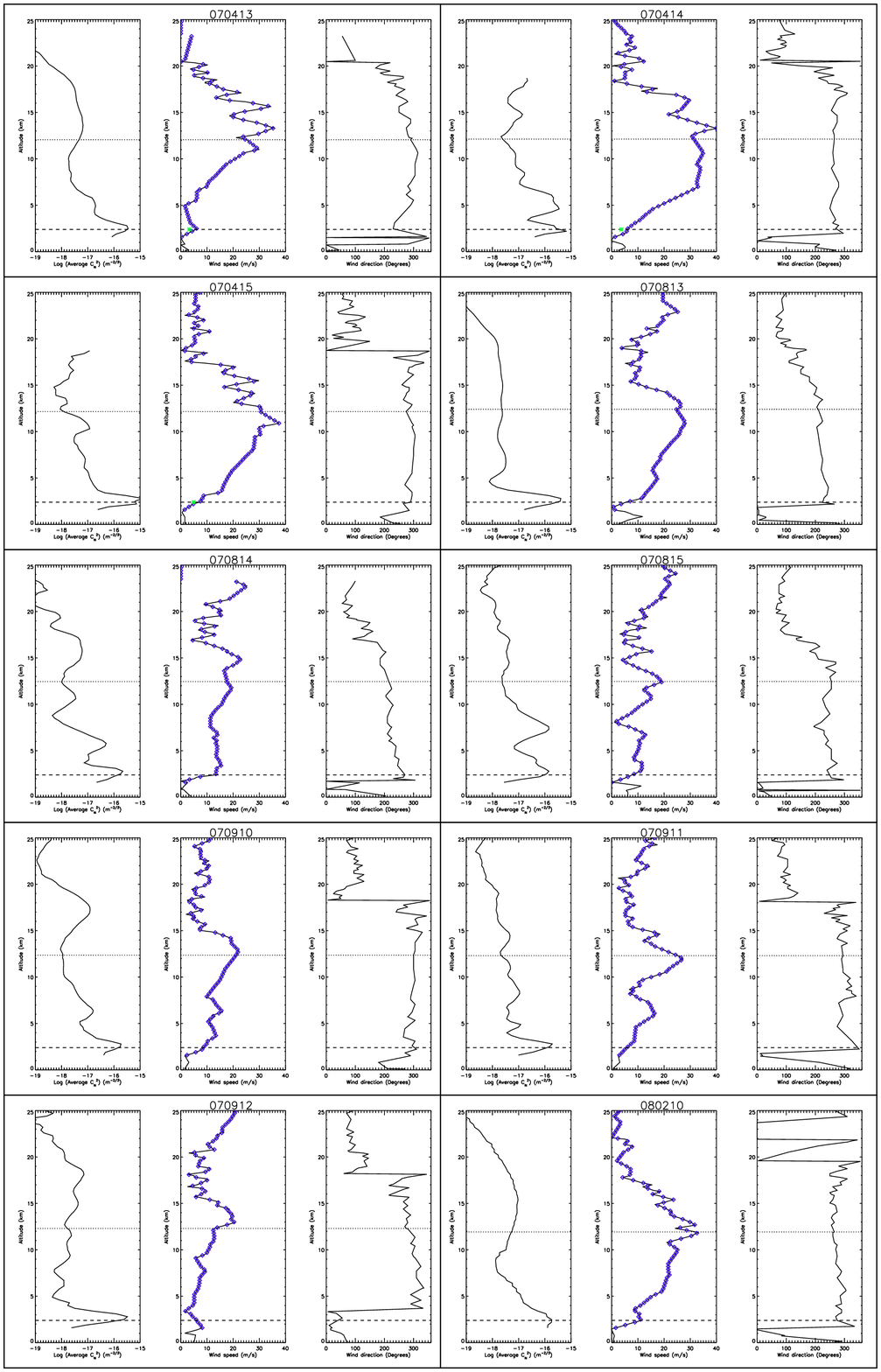}
 \caption{Continuation. }
\label{todos_perfiles1}
\end{figure*}

\begin{figure*}
\centering
\addtocounter{figure}{-1} 
\includegraphics[scale=0.70]{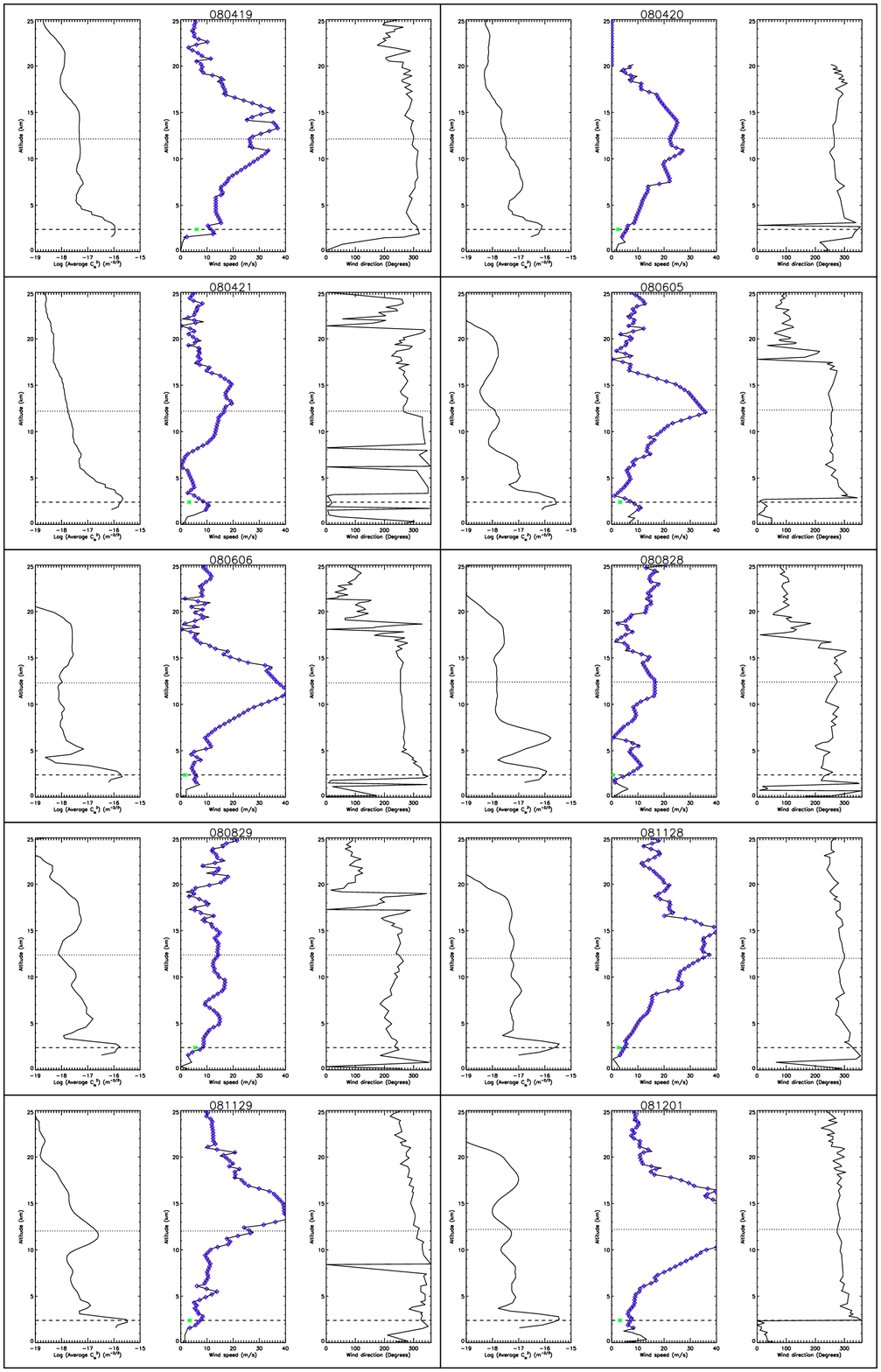}
 \caption{Continuation. }
\label{todos_perfiles}
\end{figure*}

\newpage

\section{APPENDIX B}

This appendix includes the dates and wind measurements of 100 nights distributed
 from 2003 to 2008 used to study the connection of the average velocity of the
 turbulence and high altitude winds at the Teide observatory. We have also
 included the number of individual profiles obtained from G-SCIDAR observations
  approximately during the ballon ascent period used to derive the average
 C$_N^2$ profiles. We also list the computed AO parameters from the C$_N^2$(h) and V(h) profiles.

\begin{table*}
\centering
\caption{Number of individual turbulence profiles from G-SCIDAR measurement
 used to derived the mean C$_N^2$(h) profile corresponding aproximately to the balloon ascent period during 2003, ground level wind (V$_{ground}$), wind at the 200 mbar pressure level (V$_{200}$), peak value at the wind vertical profile, and the derived AO parameters (Fried's parameter, isoplanatic angle, average velocity of the turbulence, coherence time, and turbulence characteristic altitude). The Fried's parameter and isoplanatic angle correspond to the average value derived from G-SCIDAR measurements during the corresponding balloon ascent and their uncertanties only indicate the standard deviation of the individual values.}
\begin{tabular}{|c|c|c|c|c|c|c|c|c|c|}\hline
{\bf Date} & {\bf Number} & V$_{ground}$ & V$_{200}$ & V$_{max}$ & {\it r}$_0\pm\sigma_{r_0}$ & $\theta_0\pm\sigma_{\theta_0}$ & V$_0$ & $\tau_0$ & $\hat{H}$\\
&  {\bf of profiles} & ( ms$^{-1}$ ) & ( ms$^{-1}$ ) & ( ms$^{-1}$ ) & ( cm ) & ( arcsec ) & ( m/s ) & ( ms ) & ( km ) \\\hline

2003-02-06 & 100 &     7.700 &  33.953 &  37.040 &  15.032$\pm$2.202 &   3.476$\pm$1.142 & 10.710  &   4.407 & 2.801  \\
2003-02-09 & 164 &     9.889 &  13.890 &  22.636 &  12.274$\pm$1.328 &   2.468$\pm$0.521 & 15.346  &   2.511 & 3.221  \\
2003-04-23 & 104 &     3.950 &  29.323 &  35.497 &  15.373$\pm$3.443 &   7.793$\pm$5.269 &  6.935  &   6.960 & 1.277  \\
2003-05-22 & 101 &     1.458 &  27.780 &  32.924 &  20.067$\pm$2.459 &   2.388$\pm$0.467 &  9.571  &   6.583 & 5.442  \\
2003-05-23 & 198 &     1.603 &  23.150 &  46.814 &  14.826$\pm$2.261 &   2.121$\pm$0.481 &  8.869  &   5.248 & 4.527  \\
2003-05-24 & 207 &     4.921 &   7.717 &  23.664 &  16.860$\pm$1.701 &   2.650$\pm$0.558 & 11.705  &   4.523 & 4.121  \\
2003-06-06 & 100 &     3.264 &  10.803 &  18.520 &  18.201$\pm$2.024 &   6.212$\pm$2.051 &  6.175  &   9.254 & 1.897  \\
2003-06-09 & 250 &     3.009 &   3.601 &  17.491 &  19.061$\pm$2.809 &   3.559$\pm$0.859 &  4.569  &  13.098 & 3.468  \\
2003-07-22 & 258 &     0.900 &  14.919 &  27.780 &  19.443$\pm$3.485 &   2.099$\pm$0.497 &  8.626  &   7.077 & 5.998  \\
2003-07-23 & 234 &     3.194 &  12.347 &  24.693 &  20.683$\pm$2.501 &   2.560$\pm$0.522 &  5.235  &  12.405 & 5.233  \\
2003-08-04 & 255 &     2.731 &   7.717 &  26.751 &  11.350$\pm$1.084 &   2.386$\pm$0.321 & 11.202  &   3.181 & 3.081  \\
2003-08-05 & 109 &     4.444 &   6.173 &  30.867 &  13.189$\pm$1.575 &   3.167$\pm$0.449 & 13.595  &   3.055 & 2.697  \\
2003-08-06 & 263 &     3.924 &   6.688 &  29.838 &  16.704$\pm$1.789 &   2.855$\pm$0.846 &  4.844  &  10.827 & 3.790  \\
2003-08-07 & 245 &     1.500 &   8.746 &  27.266 &  19.188$\pm$2.330 &   3.186$\pm$0.430 &  4.753  &  12.675 & 3.900  \\
2003-08-28 & 270 &     2.611 &  19.034 &  27.266 &  19.428$\pm$2.714 &   4.589$\pm$1.019 &  7.760  &   7.861 & 2.741  \\
2003-08-29 & 102 &     1.806 &  28.294 &  31.381 &  19.208$\pm$3.006 &   1.774$\pm$0.355 &  8.661  &   6.963 & 7.012  \\
2003-08-30 & 235 &     4.667 &  31.896 &  34.468 &  15.794$\pm$1.631 &   2.363$\pm$0.457 & 14.679  &   3.378 & 4.328  \\
2003-08-31 & 273 &     3.778 &  26.237 &  29.323 &  12.507$\pm$1.220 &   1.690$\pm$0.337 & 11.642  &   3.373 & 4.793  \\
2003-09-13 & 214 &     1.852 &  28.294 &  30.352 &  16.157$\pm$2.815 &   3.195$\pm$0.670 &  6.950  &   7.300 & 3.275  \\
2003-09-15 & 258 &     2.472 &  27.780 &  31.381 &  14.027$\pm$1.267 &   2.070$\pm$0.299 &  8.675  &   5.077 & 4.387  \\ \hline
\end{tabular}
\label{data03}
\end{table*}

\begin{table*}
\centering
\caption{Same as table \ref{data03} but for 2004.}
\begin{tabular}{|c|c|c|c|c|c|c|c|c|c|c|c|}\hline
{\bf Date} & {\bf Number} & V$_{ground}$ & V$_{200}$ & V$_{max}$ & {\it r}$_0\pm\sigma_{r_0}$ & $\theta_0\pm\sigma_{\theta_0}$ & V$_0$ & $\tau_0$ & $\hat{H}$ \\
&  {\bf of profiles} & ( ms$^{-1}$ ) & ( ms$^{-1}$ ) & ( ms$^{-1}$ ) & ( cm ) & ( arcsec ) & ( m/s ) & ( ms ) & ( km )\\\hline

2004-02-07  &  207   &  5.347 &  65.334 &  65.334 &  17.070$\pm$1.924 &   2.596$\pm$0.934 & 16.177 &  3.313 &  4.258  \\
2004-03-02  &  154   &  6.032 &  18.006 &  24.179 &  11.716$\pm$1.575 &   5.614$\pm$1.615 & 14.872 &  2.473 &  1.351  \\
2004-03-03  &  293   &  6.903 &  26.237 &  27.780 &  11.068$\pm$1.219 &   4.679$\pm$1.450 & 10.527 &  3.301 &  1.531  \\
2004-04-22  &  155   &  2.878 &  26.751 &  41.670 &  12.772$\pm$1.450 &   2.946$\pm$0.728 &  8.497 &  4.719 &  2.807  \\
2004-04-23  &  305   &  4.398 &  42.699 &  43.213 &  11.960$\pm$0.725 &   2.628$\pm$0.531 & 11.273 &  3.331 &  2.947  \\
2004-04-24  &  227   &  3.056 &  41.156 &  48.872 &  15.284$\pm$2.145 &   2.099$\pm$0.393 & 13.475 &  3.561 &  4.716  \\
2004-05-28  &  204   &  4.722 &  15.433 &  23.150 &  23.625$\pm$3.148 &   3.004$\pm$0.517 &  7.280 & 10.189 &  5.094  \\
2004-05-29  &  307   &  5.625 &   8.746 &  22.636 &  14.782$\pm$1.324 &   2.221$\pm$0.334 &  7.586 &  6.118 &  4.310  \\
2004-05-30  &  223   &  2.996 &   8.231 &  21.092 &  18.649$\pm$2.091 &   5.673$\pm$1.797 &  3.271 & 17.902 &  2.128  \\
2004-05-31  &  309   &  6.343 &   7.202 &  23.664 &  10.160$\pm$1.023 &   2.573$\pm$0.623 &  7.176 &  4.444 &  2.557  \\
2004-07-17  &  273   &  3.843 &  14.919 &  32.924 &  14.305$\pm$2.106 &   2.098$\pm$0.600 & 10.429 &  4.307 &  4.415  \\
2004-07-18  &  223   &  6.490 &  26.751 &  27.780 &  11.642$\pm$1.442 &   2.117$\pm$0.489 & 13.240 &  2.761 &  3.561  \\
2004-08-14  &   22   &  4.167 &  12.347 &  26.237 &  14.960$\pm$1.039 &   3.041$\pm$0.410 &  9.985 &  4.704 &  3.186  \\
2004-08-15  &  201   &  2.917 &  10.289 &  25.722 &  11.609$\pm$1.833 &   2.757$\pm$0.473 &  8.951 &  4.072 &  2.727  \\
2004-08-16  &  158   &  3.750 &  16.977 &  23.150 &  20.123$\pm$4.656 &   2.491$\pm$0.706 & 10.043 &  6.291 &  5.232  \\
2004-09-03  &  180   &  2.870 &   5.659 &  32.410 &  19.875$\pm$3.246 &   2.111$\pm$0.424 &  4.872 & 12.809 &  6.099  \\
2004-09-04  &  167   &  2.870 &   8.231 &  24.179 &  14.814$\pm$1.508 &   2.118$\pm$0.566 &  5.186 &  8.968 &  4.529  \\
2004-09-05  &   22   &  4.086 &  14.404 &  26.751 &  10.227$\pm$1.358 &   1.728$\pm$0.182 &  8.834 &  3.635 &  3.833  \\
2004-09-06  &  234   &  4.683 &  16.462 &  27.266 &  13.068$\pm$2.486 &   1.707$\pm$0.462 &  9.562 &  4.291 &  4.956  \\
2004-10-23  &  206   &  4.630 &   4.630 &  18.520 &  14.305$\pm$1.173 &   2.225$\pm$0.284 &  9.028 &  4.975 &  4.164  \\
2004-10-24  &  213   &  3.121 &  14.404 &  19.549 &  12.200$\pm$1.222 &   1.851$\pm$0.491 &  7.784 &  4.921 &  4.268  \\ \hline
\end{tabular}
\label{data04}
\end{table*}

\begin{table*}
\centering
\caption{Same as table \ref{data03} but for 2005.}
\begin{tabular}{|c|c|c|c|c|c|c|c|c|c|c|}\hline
{\bf Date} & {\bf Number} & V$_{ground}$ & V$_{200}$ & V$_{max}$ & {\it r}$_0\pm\sigma_{r_0}$ & $\theta_0\pm\sigma_{\theta_0}$ & V$_0$ & $\tau_0$ & $\hat{H}$ \\
&  {\bf of profiles} & ( ms$^{-1}$ ) & ( ms$^{-1}$ ) & ( ms$^{-1}$ ) & ( cm ) & ( arcsec ) & ( m/s ) & ( ms ) & ( km ) \\\hline
2005-03-17  &   38    &     8.889 &  26.237 &  36.526 &  13.146$\pm$1.378  &     1.626$\pm$0.197 & 12.454 &  3.314 &  5.236  \\
2005-03-19  &   41    &     6.528 &  35.497 &  54.531 &  14.059$\pm$3.931  &     1.537$\pm$0.243 &  7.360 &  5.998 &  5.923  \\
2005-04-18  &   82    &     2.083 &  14.919 &  30.867 &  18.247$\pm$2.233  &     2.085$\pm$0.197 &  9.493 &  6.035 &  5.668  \\
2005-04-20  &   55    &     3.681 &  19.034 &  28.294 &  20.417$\pm$2.512  &     2.179$\pm$0.237 &  9.232 &  6.943 &  6.067  \\
2005-05-06  &  338    &     3.843 &  32.410 &  37.554 &  13.560$\pm$1.104  &     2.360$\pm$0.454 &  9.701 &  4.388 &  3.722  \\
2005-05-07  &  349    &     6.019 &  46.300 &  46.814 &  12.786$\pm$2.018  &     1.981$\pm$0.275 & 16.413 &  2.446 &  4.180  \\
2005-05-29  &  314    &     4.564 &  29.838 &  29.838 &  19.855$\pm$4.514  &     3.243$\pm$0.559 &  8.736 &  7.136 &  3.964  \\
2005-05-30  &  307    &     3.556 &  31.896 &  31.896 &  15.108$\pm$2.309  &     2.498$\pm$1.019 & 11.260 &  4.213 &  3.916  \\
2005-06-08  &  242    &     3.889 &  20.578 &  24.179 &  14.723$\pm$3.359  &     4.242$\pm$1.592 &  6.236 &  7.413 &  2.248  \\
2005-06-09  &  241    &     4.615 &  23.664 &  27.266 &  15.275$\pm$3.688  &     5.074$\pm$2.139 &  8.305 &  5.775 &  1.949  \\
2005-07-09  &   48    &     5.500 &  13.376 &  27.780 &  12.193$\pm$0.879  &     2.454$\pm$0.231 &  5.540 &  6.910 &  3.217  \\
2005-07-10  &  293    &     5.000 &  11.318 &  31.381 &   9.424$\pm$0.710  &     3.117$\pm$0.656 &  9.249 &  3.200 &  1.957  \\
2005-07-11  &  336    &     5.444 &  11.318 &  31.381 &  13.385$\pm$1.275  &     2.550$\pm$0.478 & 10.729 &  3.917 &  3.399  \\
2005-08-04  &  111    &     2.130 &  11.318 &  29.838 &  11.786$\pm$1.395  &     3.843$\pm$0.861 &  5.337 &  6.933 &  1.986  \\
2005-08-06  &  357    &     4.167 &  12.861 &  29.838 &  13.074$\pm$1.638  &     1.577$\pm$0.301 &  9.922 &  4.137 &  5.369  \\
2005-08-07  &  346    &     3.571 &  14.404 &  33.439 &  12.369$\pm$1.192  &     2.318$\pm$0.576 &  9.445 &  4.111 &  3.455  \\
2005-08-28  &  206    &     7.303 &  23.150 &  26.751 &  11.982$\pm$1.303  &     2.310$\pm$0.366 & 15.104 &  2.491 &  3.358  \\
2005-08-29  &  251    &     4.675 &  24.693 &  27.266 &  12.053$\pm$1.282  &     1.514$\pm$0.296 & 12.303 &  3.076 &  5.156  \\
2005-09-01  &  146    &     4.802 &   7.717 &  24.179 &  17.972$\pm$2.087  &     2.285$\pm$0.380 &  5.234 & 10.781 &  5.093  \\
2005-09-02  &  275    &     4.944 &   8.746 &  26.237 &  14.426$\pm$1.416  &     2.254$\pm$0.531 &  5.985 &  7.568 &  4.145  \\
2005-09-03  &  195    &     4.792 &  27.780 &  34.468 &  13.936$\pm$2.054  &     1.537$\pm$0.363 & 10.157 &  4.308 &  5.870  \\
2005-09-25  &  323    &     4.861 &  27.266 &  34.468 &  11.418$\pm$1.347  &     2.138$\pm$0.488 &  8.711 &  4.116 &  3.459  \\
2005-09-26  &  323    &     3.426 &  13.376 &  20.578 &  13.645$\pm$1.911  &     2.670$\pm$0.506 &  4.041 & 10.602 &  3.309  \\ \hline
\end{tabular}
\label{data05}
\end{table*}

\begin{table*}
\centering
\caption{Same as table \ref{data03} but for 2006.}
\begin{tabular}{|c|c|c|c|c|c|c|c|c|c|c|}\hline
{\bf Date} & {\bf Number} & V$_{ground}$ & V$_{200}$ & V$_{max}$ & {\it r}$_0\pm\sigma_{r_0}$ & $\theta_0\pm\sigma_{\theta_0}$ & V$_0$ & $\tau_0$ & $\hat{H}$ \\
&  {\bf of profiles} & ( ms$^{-1}$ ) & ( ms$^{-1}$ ) & ( ms$^{-1}$ ) & ( cm ) & ( arcsec ) & ( m/s ) & ( ms ) & ( km )\\\hline
2006-03-27 &    168   &      2.738 &  19.034 &  28.294 &  21.165$\pm$4.225 &   2.629$\pm$0.455 & 10.637 &  6.247 &  5.215  \\
2006-03-29 &    304   &      2.963 &  30.867 &  31.381 &  11.968$\pm$1.669 &   2.883$\pm$0.591 &  7.320 &  5.133 &  2.688  \\
2006-04-07 &    143   &      6.146 &  20.063 &  31.896 &  12.097$\pm$1.968 &   4.569$\pm$1.318 & 10.733 &  3.539 &  1.714  \\
2006-04-25 &    339   &      4.907 &  10.289 &  15.948 &  17.625$\pm$2.964 &   5.140$\pm$1.698 &  7.679 &  7.206 &  2.220  \\
2006-04-26 &    344   &      5.417 &  19.034 &  24.179 &  18.785$\pm$2.705 &   4.210$\pm$1.320 &  9.022 &  6.537 &  2.889  \\
2006-04-27 &    329   &      5.694 &  18.006 &  25.722 &  12.409$\pm$1.271 &   1.968$\pm$0.380 &  9.441 &  4.127 &  4.083  \\
2006-05-26 &    177   &      5.556 &  30.352 &  36.526 &   9.927$\pm$0.953 &   2.023$\pm$0.734 & 10.105 &  3.084 &  3.177  \\
2006-05-28 &    288   &      4.861 &  26.751 &  30.867 &  13.183$\pm$1.987 &   4.449$\pm$1.843 &  7.761 &  5.333 &  1.919  \\
2006-06-08 &    194   &      6.065 &  49.387 &  50.930 &  13.671$\pm$0.998 &   1.611$\pm$0.307 & 16.817 &  2.552 &  5.495  \\
2006-07-23 &    224   &      2.222 &  12.861 &  26.237 &  17.603$\pm$3.545 &   2.897$\pm$0.464 &  4.136 & 13.364 &  3.935  \\
2006-07-24 &     51   &      2.400 &  11.832 &  30.867 &  23.963$\pm$3.530 &   3.121$\pm$0.651 &  5.263 & 14.295 &  4.972  \\ \hline
\end{tabular}							
\label{data06}							
\end{table*}

\begin{table*}							
\centering							
\caption{Same as table \ref{data03} but for 2007.}					
\begin{tabular}{|c|c|c|c|c|c|c|c|c|c|c|}\hline
{\bf Date} & {\bf Number} & V$_{ground}$ & V$_{200}$ & V$_{max}$ & {\it r}$_0$ & $\sigma_{r_0}\pm\theta_0\pm\sigma_{\theta_0}$ & V$_0$ & $\tau_0$ & $\hat{H}$ \\
&  {\bf of profiles} & ( ms$^{-1}$ ) & ( ms$^{-1}$ ) & ( ms$^{-1}$ ) & ( cm ) & ( arcsec ) & ( m/s ) & ( ms ) & ( km ) \\\hline
2007-03-01  &   128    &    5.764 &   5.144 &  30.867 &  18.445$\pm$2.542 &   3.075$\pm$1.451 &  8.743 &  6.624 &   3.885   \\
2007-03-02  &   155    &    3.843 &  11.318 &  19.034 &  14.801$\pm$1.292 &   2.414$\pm$0.544 &  6.900 &  6.734 &   3.970   \\
2007-03-04  &   245    &    3.194 &  11.832 &  25.208 &  19.942$\pm$3.325 &   1.714$\pm$0.484 & 13.181 &  4.750 &   7.535   \\
2007-03-31  &    86    &    1.597 &  40.641 &  45.271 &  15.603$\pm$1.536 &   1.893$\pm$0.413 & 18.485 &  2.650 &   5.338   \\
2007-04-01  &   137    &    4.944 &  39.612 &  40.641 &  15.544$\pm$2.363 &   1.682$\pm$0.326 & 18.524 &  2.635 &   5.983   \\
2007-04-13  &    90    &    3.333 &  25.208 &  35.497 &  12.837$\pm$0.943 &   2.820$\pm$0.621 &  8.111 &  4.969 &   2.948   \\
2007-04-14  &   179    &    5.704 &  31.381 &  40.127 &   7.991$\pm$1.454 &   1.427$\pm$0.266 & 16.947 &  1.481 &   3.627   \\
2007-04-15  &    89    &    6.907 &  30.352 &  37.554 &   8.420$\pm$1.422 &   3.353$\pm$1.058 &  9.819 &  2.693 &   1.626   \\
2007-08-13  &   320    &    7.500 &  24.693 &  28.294 &  14.383$\pm$3.656 &   4.315$\pm$2.413 & 10.530 &  4.289 &   2.158   \\
2007-08-14  &   224    &    7.000 &  17.491 &  25.208 &  16.728$\pm$2.252 &   3.047$\pm$0.528 & 13.833 &  3.797 &   3.555   \\
2007-08-15  &    37    &    6.500 &  19.034 &  26.751 &  13.417$\pm$1.948 &   1.715$\pm$0.154 &  9.401 &  4.481 &   5.065   \\
2007-09-10  &   322    &    6.800 &  21.092 &  22.121 &  17.910$\pm$1.929 &   2.494$\pm$0.663 & 10.044 &  5.599 &   4.651   \\
2007-09-11  &   250    &    6.000 &  25.208 &  27.266 &  20.117$\pm$2.376 &   2.998$\pm$0.402 &  9.232 &  6.842 &   4.346   \\
2007-09-12  &    24    &    5.500 &  12.861 &  26.237 &  17.742$\pm$1.265 &   2.743$\pm$0.270 &  5.982 &  9.312 &   4.189   \\ \hline
\end{tabular}						       		
\label{data07}						       		
\end{table*}						       		
							       		
\begin{table*}
\centering
\caption{Same as table \ref{data03} but for 2008.}
\begin{tabular}{|c|c|c|c|c|c|c|c|c|c|c|}\hline
{\bf Date} & {\bf Number} & V$_{ground}$ & V$_{200}$ & V$_{max}$ & {\it r}$_0\pm\sigma_{r_0}$ & $\theta_0\pm\sigma_{\theta_0}$ & V$_0$ & $\tau_0$ & $\hat{H}$ \\
&  {\bf of profiles} & ( ms$^{-1}$ ) & ( ms$^{-1}$ ) & ( ms$^{-1}$ ) & ( cm ) & ( arcsec ) & ( m/s ) & ( ms ) & ( km )\\\hline
2008-02-10 &  58   &     6.151 &  32.410 &  32.924 &  15.075$\pm$1.295 &   2.081$\pm$0.354 & 13.277 &  3.575 &  4.691  \\
2008-04-19 & 171   &     5.556 &  26.237 &  37.040 &  18.404$\pm$1.671 &   2.743$\pm$0.372 & 15.590 &  3.707 &  4.345  \\
2008-04-20 & 188   &     2.381 &  22.121 &  27.266 &  23.052$\pm$4.316 &   3.156$\pm$0.446 & 12.359 &  5.857 &  4.731  \\
2008-04-21 &  50   &     3.214 &  16.462 &  20.063 &  13.705$\pm$1.630 &   4.162$\pm$0.859 &  6.895 &  6.241 &  2.132  \\
2008-06-05 & 265   &     3.194 &  34.982 &  36.011 &  15.428$\pm$2.558 &   4.915$\pm$1.562 &  6.795 &  7.129 &  2.033  \\
2008-06-06 &  97   &     1.667 &  37.040 &  41.156 &  19.875$\pm$1.899 &   4.412$\pm$1.039 &  7.296 &  8.553 &  2.917  \\
2008-08-28 & 310   &     0.347 &  16.462 &  27.266 &  15.443$\pm$2.180 &   2.579$\pm$0.518 &  6.370 &  7.612 &  3.878  \\
2008-08-29 & 329   &     5.444 &  14.404 &  25.722 &  19.818$\pm$2.726 &   2.980$\pm$0.566 &  9.874 &  6.302 &  4.307  \\
2008-11-28 &  75   &     2.639 &  34.468 &  42.699 &  14.762$\pm$1.186 &   2.377$\pm$0.377 & 12.282 &  3.774 &  4.022  \\
2008-11-29 & 313   &     3.444 &  26.751 &  41.670 &  15.667$\pm$1.866 &   2.827$\pm$0.659 & 12.414 &  3.963 &  3.588  \\
2008-12-01 &  89   &     3.125 &  56.074 &  57.618 &  14.485$\pm$1.632 &   2.342$\pm$0.395 & 13.219 &  3.441 &  4.005  \\ \hline

\end{tabular}
\label{data08}
\end{table*}

\end{document}